\documentstyle[prd,aps,epsfig]{revtex}
\begin{document}
\title{The superfluid phases of quark matter: Ginzburg-Landau theory and color
neutrality}
\author{Kei Iida$^{1,2}$ and Gordon Baym$^{1}$}
\address{$^{1}$Department of Physics, University of Illinois at
Urbana-Champaign \\ 1110 West Green Street, Urbana, Illinois 61801-3080}
\address{$^{2}$Department of Physics, University of Tokyo
\\ 7-3-1 Hongo, Bunkyo, Tokyo 113-0033, Japan}
\date{\today}
\maketitle
\begin{abstract}

    We systematically apply Ginzburg-Landau theory to determine BCS pairing
in a strongly-coupled uniform superfluid of three-flavor massless quarks in
flavor equilibrium.  We elucidate the phase diagram near the critical 
temperature in the space of the parameters characterizing the 
thermodynamic-potential terms of fourth order in the pairing gap.  Within the
color and flavor antisymmetric channel with zero total angular momentum, the
phase diagram contains an isoscalar, color-antitriplet phase and a 
color-flavor-locked phase, reached by a second order phase transition from the
normal state, as well as states reached by a first order phase transition.  We
complement the general Ginzburg-Landau approach by deriving the high-density 
asymptotic form of the Ginzburg-Landau free energy from the finite temperature
weak-coupling gap equation.  The dynamically-screened, long-range color 
magnetic interactions are explicitly taken into account in solving the gap 
equation.  We find that in the limit of weak coupling, the isoscalar, 
color-antitriplet phase has higher free energy near the transition temperature
than the color-flavor locked phase.

    In view of the fact that deconfined quark matter must be color charge
neutral, we incorporate the constraint of overall color neutrality into the
general Ginzburg-Landau theory and the gap equation.  This constraint yields a
disparity in the chemical potential between colors and reduces the size of the
pairing gap, in the presence of the anisotropy of the order parameters in
color space.  In comparison with the case in which there are no chemical
potential differences between colors and hence the superfluid generally has
nonzero net color charge, we find that while the constraint of color
neutrality has only negligible effects on the gap in the weak coupling regime,
it appreciably destabilizes the isoscalar, color-antitriplet phase in the
strong coupling regime without affecting the color-flavor-locked phase.

\end{abstract}

\section{Introduction}
\label{sec:intro}

    The possibility that degenerate relativistic quark matter becomes a color
superconductor at low temperatures has been considered for the past two
decades.  As the seminal papers \cite{barrois,BL} noted, the quark-quark
interaction in the color antitriplet channel is attractive and drives a Cooper
pairing instability in the system, even in the limit of high densities where
the Fermi energy of the quarks dominates over the one-gluon exchange
interaction energy.  Work in the intervening period has concentrated on the
mean field approach, strictly valid only in the weak coupling limit (see,
e.g., Refs.\ \cite{RP} and \cite{ABR} for reviews).  As the density is
lowered, nonperturbative effects arising from self-couplings of the gluon
field prevail, finally leading to a confinement transition into hadronic
matter.  Strong coupling effects can modify the equilibrium order parameter
from the weak coupling prediction, as effects of spin fluctuation exchange do
in superfluid $^{3}$He \cite{helium}.  The resultant change in the
color-superconducting phase can also affect the breaking of chiral symmetry
\cite{wilczek,rajagopal,benoit}.  The properties of the color
superconductivity such as the pairing gap and the critical temperature, $T_c$,
have yet to be derived in the strong coupling regime in a rigorous way.

    In this paper we examine color superconductors by means of a general
Ginzburg-Landau approach, which permits us to determine the coarse grained
features of such systems at temperatures just below $T_c$, for arbitrary QCD
coupling constant, $g$.  This approach, pioneered by Anderson and Brinkman
\cite{AB} and Mermin and Stare \cite{MS} in the context of superfluid
$^{3}$He, reveals the most energetically favorable phase just below $T_c$, in
terms of the parameters characterizing the thermodynamic potential to fourth
order in the pairing gap.  The first application of the general
Ginzburg-Landau theory was made by Bailin and Love \cite{BL} to quark pairing
with one flavor and total angular momentum $J=1$.  Here we shall consider more
general Cooper pairing between $u$, $d$, and $s$ quarks with $J=0$.  We
complement this approach by deriving the parameters controlling the fourth
order terms from the weak coupling gap equation.  Previous work
\cite{tom,EHHS} has systematically investigated such pairing at zero 
temperature; the present study provides a systematic elucidation of the 
equilibrium properties near $T_{c}$.

    The zero-temperature pairing gap and hence $T_{c}$ is predicted to be
$\sim$ 10--100 MeV for baryon chemical potentials $\sim$ 1 GeV
\cite{instanton}.  This prediction relies on extrapolation from weak coupling
into the low density, nonperturbative regime, by incorporating into the BCS
gap equations effective interactions modeled after instanton-mediated
interactions, in such a way as to reproduce constituent quark masses.
Application of the Wilson renormalization group to analysis of the stability
of a Fermi liquid against Cooper pairing \cite{RG} suggests that four-fermion
couplings, induced between two particles with zero total momentum by one-gluon
exchange or by instantons, grow logarithmically with momentum as higher modes
are successively integrated out closer to the Fermi surface; the scattering
amplitudes eventually reach a singularity -- a Landau pole -- in a way
dependent on the number of flavors involved in the pairing.  For two flavors
($u$, $d$), both gluon and instanton-induced interactions play a role in
opening an energy gap in the isoscalar, color-antitriplet channel with zero
total angular momentum ($J=0$).  Pairing in this channel partially breaks 
baryon number symmetry and global color rotational invariance, but restores 
chiral symmetry.  For three flavors ($u$, $d$, $s$), on the other hand, a 
color-flavor locked state \cite{ARW} arises in the $J=0$ channel mainly from 
the gluon-induced interactions when the strange quark mass is sufficiently 
small.  This state is invariant under simultaneous exchange of color and 
flavor, but not under single exchange of color or flavor; chiral symmetry as 
well as baryon number symmetry and global color rotational invariance are 
broken.  In the high density regime where the interactions are dominated by 
one-gluon exchange, the color magnetic (transverse) force, which is screened 
only dynamically by Landau damping of its mediators \cite{BMPR,art}, is
sufficiently long ranged to alter the dependence of the pairing gap on the QCD
coupling constant $g$ from the BCS result.  This fact was first noted by Son
\cite{son} using a renormalization group method and an approximate solution to
the relevant gap equation.

    As in prior papers, we focus on the equilibrium properties of an
ultrarelativistic color superconductor that is {\it homogeneous}, in the sense
that the real gluon field vanishes everywhere and the order parameter is
everywhere continuous in magnitude and orientation; we denote such a state
as {\it superfluid quark matter}.  Such homogeneity is similar to that in
superfluid $^{3}$He and superfluid neutron matter, as noted by Bailin and Love
\cite{BL}, because in both cases breaking of the global $U(1)$ gauge symmetry
is accompanied by global symmetry breaking associated with the internal
degrees of freedom.  In superfluid quark matter, the possible order parameters
are generally {\it anisotropic} in color space (see, e.g., Ref.\ \cite{BL}), a
situation analogous to superfluid $^{3}$He in which, as seen experimentally,
the anisotropy lies in spin space \cite{helium}.

    We first restrict ourselves to the case, normally assumed in earlier
investigations, in which the chemical potentials of different color and flavor
quarks are equal.  We obtain the thermodynamic potential difference between
the superfluid and normal phases near $T_{c}$ from the Ginzburg-Landau
approach.  The terms of second and fourth order in the pairing gap are
constrained by invariances of the grand canonical Hamiltonian and by the
structure of the order parameters, assumed here to be antisymmetric in color
and flavor space.  We then identify the degenerate sets of order parameters
corresponding to local energy minima as isoscalar color-antitriplet and
color-flavor locked states, and determine their condensation energies.  We
draw the resultant equilibrium phase diagram near $T_{c}$ in the space of
parameters characterizing the fourth order terms.  We find that in the limit
of weak coupling, the isoscalar, color-antitriplet phase is less favorable
than the color-flavor locked phase near $T_{c}$.

    In general, determination of the Ginzburg-Landau parameters requires
inclusion of strong coupling effects.  In the weak coupling limit, we anchor
the general Ginzburg-Landau approach by deriving the parameters from the
relevant weak coupling gap equations at finite temperature \cite{BL,PR1,PR3},
including the infrared structure of the gluon propagator as in Ref.\
\cite{PR3}.  The latter behavior determines the $g$ dependence of the pairing
gap in the weak coupling limit.  Debye screening of the electric gluons and
Landau damping of the magnetic gluons, calculated with one-gluon exchange as
modified by a normal medium, provide effective infrared cutoffs of the
respective scattering amplitudes.

    Motivated by the fact that deconfined quark matter must be in an overall
color singlet state, we also consider the equilibrium properties of superfluid
quark matter under the condition that it is color-charge neutral.  In normal
quark matter, in which global color rotational invariance is ensured by QCD
interactions, the constraint of color neutrality leads to equality of the
chemical potentials for quarks of different colors (see, e.g., Ref.\
\cite{FM}).  Superfluid quark matter, on the other hand, generally has a
preferred direction in color space.  Such violation of color rotational
invariance together with the requirement of overall color neutrality leads to
differences in the chemical potential between colors, in a way dependent on
the degree of anisotropy of the order parameters in color space.  This
disparity in chemical potentials in turn alters the pairing gap, a feature
seen in the general Ginzburg-Landau analysis.  Beyond the usual terms of
second and fourth order in the pairing gap, color neutrality adds terms
dependent on the chemical potential differences between colors.  Such extra
terms, which act in the isoscalar, color-antitriplet channel, do not remove
the degeneracy of the order-parameter sets occurring for equal color chemical
potentials and renormalize the coefficients of the fourth order terms so as to
reduce the pairing gap.  In order to estimate the influence of color
neutrality on the gap in the weak coupling limit, we incorporate color
chemical potential differences in the gap equation, and find that in the weak
coupling limit, such differences do not significantly modify the gap.
Although it does not affect the color-flavor-locked phase, the constraint of
color neutrality can appreciably raise the free energy of the isoscalar,
color-antitriplet phase in the strong coupling regime.

    We base our calculations in the weak coupling limit on the
field-theoretical description of an ordinary superconductor as advanced by
Nambu \cite{nambu} and Eliashberg \cite{eliashberg}, who reformulated the
original BCS problem in terms of electron-phonon interactions.  Such a
formalism is useful when the interactions are nonlocal in time, as are the
color magnetic interactions in the relativistic plasma.  This tool was
generalized to a relativistic regime by others in earlier investigations
(e.g., Ref.\ \cite{BL}).

    We focus in this paper on a system in flavor equilibrium, with no chemical
potential differences between flavors (as long as the electrical charge of the
quarks is neglected).  As a consequence, the energy of the superfluid
is invariant under rotations of the equilibrium states in flavor space.
The influence of the possible flavor chemical potential differences on the
phase diagram, examined by others in earlier investigations
\cite{ABR,bedaque}, resolves this degeneracy.  Such differences may also give
rise to an inhomogeneous phase in which the pairing gap varies periodically in
ordinary space.  The conditions determining the chemical potential differences
between flavors depend, however, sensitively on the extent to which the system
is out of flavor equilibrium.

    We summarize the main conclusions of this paper.  By applying general
Ginzburg-Landau theory to the $J=0$, color and flavor antisymmetric pairing
states of superfluid quark matter, we elucidate the phase diagram near $T_c$
in the space of the parameters controlling the fourth order terms in the
pairing gap.  We find that the phase diagram contains an isoscalar
color-antitriplet phase and a color-flavor locked phase, both reached by a
second order phase transition, as well as a region of Ginzburg-Landau 
parameters for which the transition to the superfluid phase is of first order.
The detailed structure of the superfluid phase in this regime depends on the 
nature of the sixth order terms in the expansion.  In the limit of weak 
coupling, the color-flavor locked phase is more favorable than the isoscalar
color-antitriplet phase.  We also expand the general Ginzburg-Landau approach
and the Nambu-Eliashberg formalism to incorporate differences in the chemical
potential between colors, required to preserve color-charge neutrality of the
system.  We find that the color neutrality constraint in the presence of
anisotropy of the order parameters in color space tends to suppress the gap.

    Color superconductivity in quark matter becomes an astrophysically
interesting problem if neutron star interiors are sufficiently dense that they
contain quark matter cores \cite{HP}.  Generally, a quark superfluid in a
neutron star would not be electrically neutral since each quark has fractional
electric charge; rather it would coexist with electrons (and muons) in such a
way as to ensure electric neutrality in the system.  Due to the dually charged
nature of the quarks, macroscopic manifestations of both color and
electromagnetic superconductivity such as Meissner effects, generation of
London fields, and vortex formation are expected from magnetic fields and
rotations as observed in these celestial objects.  We will discuss these
issues, which may be relevant to magnetic structure, cooling, and rotational
evolution of the neutron stars \cite{ABR,NS}, elsewhere \cite{paperIII}.

    In Sec.\ \ref{sec:GGL}, we construct the generalized Ginzburg-Landau
theory including color chemical potential differences, and apply it to color
and flavor antisymmetric pairing with $J=0$.  The equilibrium phase diagrams
for various values of the parameters controlling the expansion of the free
energy are given here.  Section \ref{sec:GE} is devoted to deriving the gap
equations in the weak coupling limit, into which the color chemical potential
differences are incorporated, and to calculations of the energy gap for the
two types of optimal pairing states derived in Sec.\ \ref{sec:GGL}.  In Sec.\
\ref{sec:GL}, we calculate the thermodynamic potential difference between the
superfluid and normal phases near $T_{c}$ from the structure of the energy gap
obtained in Sec.\ \ref{sec:GE}.  Our conclusions are given in Sec.\ 
\ref{sec:conclusion}.  We use units in which $\hbar=c=k_{B}=1$.

\section{General Ginzburg-Landau approach}
\label{sec:GGL}

    In this section, we apply the general Ginzburg-Landau approach to a quark
superfluid that is in an overall color singlet state.  We then derive the
thermodynamic potentials in terms of the color chemical potential differences.
We finally draw the equilibrium phase diagrams near $T_{c}$ for the parameters
characterizing the strong-coupling effects and the color-neutrality
constraint.

    We consider uniform quark matter of temperature $T$ and baryon chemical
potential $\mu$, with number of flavors $N_{f}=3$ ($u$,$d$,$s$) and colors
$N_{c}=3$.  We neglect quark masses and electrical charges as well as chiral
condensates, and do not take into account at this stage either the color or 
electromagnetic gauge field, except for virtual gluons mediating interactions
between the quarks.  We also assume that the system has zero net color charge
and satisfies the condition for flavor equilibrium,
\begin{equation}
  \mu_{a u}=\mu_{a d}=\mu_{a s}\equiv\mu_{a}\ ,
   \label{flaveq}
\end{equation}
where $\mu_{a i}$ is the chemical potential for the quarks of color $a$
and flavor $i$.  Thus, the ensemble-averaged color charge densities vanish,
\begin{equation}
 \langle\Sigma_{i}\lambda^{\alpha}_{ab}{\bar\psi}_{ai}(x)\gamma^{0}
   \psi_{bi}(x)\rangle = 0\ ,
\end{equation}
where the $\psi_{ai}$ are the quark fields and
$\lambda^{\alpha}_{ab}$ the color generators.   As a consequence,
\begin{equation}
n_{R}=n_{G}=n_{B}\ ,
   \label{colsindia}
\end{equation}
with $n_{a}\equiv\langle\Sigma_{i}{\bar\psi}_{ai}(x)\gamma^{0}\psi_{ai}(x)
\rangle$, and
\begin{equation}
n_{ab}\equiv\langle\Sigma_{i}{\bar\psi}_{ai}(x)\gamma^{0}\psi_{bi}(x)
\rangle=0
   \label{colsinnondia}
\end{equation}
for $a\neq b$.  Condition (\ref{flaveq}), along with condition
(\ref{colsindia}), relates $\mu_{a}$ to $\mu$ as
\begin{equation}
  \mu=\sum_{a}\mu_{a}\ ;
 \label{mu}
\end{equation}
equivalently we write,
\begin{equation}
  \tilde{\mu}_{aa} \equiv \mu_{a}-\frac{\mu}{3}\ .
  \label{deltamu1}
\end{equation}
Then
\begin{equation}
\sum_{a}\tilde{\mu}_{aa}=0\ .
  \label{deltamu2}
\end{equation}
For $a\neq b$, we introduce the chemical potentials ${\tilde\mu}_{ab}$
associated with $n_{ab}$ in such a way that the grand-canonical Hamiltonian
reads $K=H-\sum_{ab}(\delta_{ab}\mu/3+{\tilde\mu}_{ab})[\sum_{i} \int
d^{3}x{\bar\psi}_{bi}(x)\gamma^{0}\psi_{ai}(x)]$.  Following Iwasaki and Iwado
\cite{iwasaki}, we can regard these ${\tilde\mu}_{ab}$ as Lagrange multipliers
that ensure condition (\ref{colsinnondia}).  The hermiticity of $K$ requires
${\tilde\mu}_{ab}={\tilde\mu}_{ba}^{*}$.

    Cooper pairing between quarks, which we assume to be in a channel with
zero total angular momentum, is most generally characterized by a
$4N_{f}N_{c}\times4N_{f}N_{c}$ gap matrix in color, flavor, and Dirac space
\cite{PR1},
\begin{eqnarray}
  \Delta(k)&=&\Delta^{(1)}(k_{0},|{\bf k}|)\gamma^{5}+
  \Delta^{(2)}(k_{0},|{\bf k}|)\mbox{\boldmath $\gamma\cdot$}{\bf {\hat k}}
  \gamma^{0}\gamma^{5}+
  \Delta^{(3)}(k_{0},|{\bf k}|)\gamma^{0}\gamma^{5}+
  \Delta^{(4)}(k_{0},|{\bf k}|)
\nonumber \\ & &
  +\Delta^{(5)}(k_{0},|{\bf k}|)
  \mbox{\boldmath $\gamma\cdot$}{\bf {\hat k}}\gamma^{0}
  +\Delta^{(6)}(k_{0},|{\bf k}|)\mbox{\boldmath $\gamma\cdot$}{\bf {\hat k}}
  +\Delta^{(7)}(k_{0},|{\bf k}|)\mbox{\boldmath $\gamma\cdot$}{\bf {\hat k}}
  \gamma^{5}+
  \Delta^{(8)}(k_{0},|{\bf k}|)\gamma^{0}\ ,
  \label{delta}
\end{eqnarray}
where $k$ is the relative four-momentum between the two quarks forming the
Cooper pair, ${\bf {\hat k}}\equiv{\bf k}/|{\bf k}|$, and
$\Delta^{(n)}_{abij}$ are the $N_{f}N_{c}\times N_{f}N_{c}$ matrices denoting
the pairing of the quark of color $a$ and flavor $i$ with that of color $b$
and flavor $j$.
 The gap $\Delta$ is formally related to the pair amplitude
$\langle \psi^{C}(x){\bar \psi}(y)\rangle$ via
\cite{nambu,eliashberg,convention}
\begin{eqnarray}
 \Delta(k)&=&ig^{2}T\sum_{n~ {\rm odd}}\int\frac{d^{3}q}{(2\pi)^{3}}
\int d^{4}(x-y)e^{iq(x-y)}\gamma^{\mu}\frac{(\lambda^{\alpha})^{T}}{2}
\nonumber \\ & & \times
\left\{\langle T[\psi^{C}(x){\bar\psi}(y)]\rangle\Gamma_{\nu\beta}^{(11)}(q,k)
+\langle T[\psi^{C}(x){\bar \psi}^{C}(y)]\rangle\Gamma_{\nu\beta}^{(21)}(q,k)
\right\}D_{\mu\nu}^{\alpha\beta}(k-q)\ ,
\end{eqnarray}
where $\psi^{C}_{ai}\equiv C{\bar \psi}^{T}_{ai}$ is the charge-conjugate
spinor ($C=i\gamma^2\gamma^0$ in the Pauli-Dirac representation),
$\Gamma^{(11)}$ is the full quark-quark-gluon vertex, $\Gamma^{(21)}$ is the
full antiquark-quark-gluon vertex, $D$ is the full gluon propagator, and the
Matsubara frequencies are given by $q_0=in\pi T$.  (In Sec.\ \ref{sec:GE} we
analyze this gap equation in the weak coupling limit; see Eq.~(\ref{wcgapeq}).)
The Pauli principle requires that $\Delta^{(n)}$ satisfy
\begin{equation}
  \Delta^{(n)}_{abij}(k_{0},|{\bf k}|)
 =\Delta^{(n)}_{baji}(-k_{0},|{\bf k}|)\ ,~~~~~n=1,...,6,
 \label{pauli1}
\end{equation}
\begin{equation}
  \Delta^{(n)}_{abij}(k_{0},|{\bf k}|)
 =-\Delta^{(n)}_{baji}(-k_{0},|{\bf k}|)\ ,~~~~~n=7,8.
 \label{pauli2}
\end{equation}

    For massless quarks the condensates are eigenstates of chirality
\cite{PR2}.  Wilson renormalization-group analyses \cite{RG} show that the
pairing instability occurs between quarks of the same chirality, rather than
between the left- and right-handed quarks.  We thus ignore the terms
$\Delta^{(3)}$, $\Delta^{(6)}$, $\Delta^{(7)}$, and $\Delta^{(8)}$ in Eq.\
(\ref{delta}), which are associated with quarks of opposite chirality
\cite{PR1}.  The $\Delta^{(4)}$ and $\Delta^{(5)}$ terms in Eq.\ (\ref{delta})
correspond to condensation in the odd-parity channel \cite{PR1}.  Effects of
instantons, which prefer even-parity to odd-parity condensates
\cite{instanton}, lead us to drop these terms.  Since only $\Delta^{(1)}$ and
$\Delta^{(2)}$ remain in Eq.\ (\ref{delta}), it is convenient to introduce
\begin{equation}
  \phi_{\pm}(k_{0},|{\bf k}|)\equiv
  \Delta^{(1)}(k_{0},|{\bf k}|)\mp\Delta^{(2)}(k_{0},|{\bf k}|)\ ,
  \label{phi}
\end{equation}
and rewrite $\Delta(k)$ as
\begin{equation}
  \Delta(k)=
  \gamma^{5}[\phi_{+}(k_{0},|{\bf k}|)\Lambda^{+}({\bf {\hat k}})
  +\phi_{-}(k_{0},|{\bf k}|)\Lambda^{-}({\bf {\hat k}})]\ .
  \label{delta1}
\end{equation}
Here, the
\begin{equation}
  \Lambda^{\pm}({\bf {\hat k}})=
  \frac{1\pm\gamma^{0}\mbox{\boldmath $\gamma\cdot$}{\bf {\hat k}}}{2}
\end{equation}
are energy projection operators for noninteracting massless quarks, and
$\phi_\pm$ denotes the quark-quark (antiquark-antiquark) pairing gap.

    We divide the difference $\Delta\Omega[\phi_{\pm}(k_{0},|{\bf k}|);
{\tilde \mu}_{ab}]$ of the thermodynamic potential near $T_{c}$ of the
superfluid phase, $\Omega_{\rm SF}$, from that of the normal phase,
$\Omega_{\rm N}$, into two parts:
\begin{equation}
 \Delta\Omega=\Omega_{0}+\Omega_{\rm CN}\ ,
 \label{deltaomega}
\end{equation}
where
\begin{equation}
  \Omega_{0}=\Delta\Omega[\phi_{\pm}(k_{0},|{\bf k}|); {\tilde
\mu}_{ab}=0]
\end{equation}
has the form of a usual Ginzburg-Landau free energy, and
\begin{equation}
  \Omega_{\rm CN}=\Delta\Omega[\phi_{\pm}(k_{0},|{\bf k}|);
  {\tilde \mu}_{ab}] -\Omega_{0}
\end{equation}
is the correction to $\Omega_{0}$ induced by the color chemical potential
differences in the superfluid (the normal phase satisfies ${\tilde
\mu}_{ab}=0$ because of rotational invariance in color space).  In expressing
$\Omega_{0}$ in terms of $\phi_{\pm}$, we follow the line of argument of
Mermin and Stare \cite{MS}.  Note that the grand-canonical Hamiltonian,
described by the chemical potential, $\mu/3$, common to all combinations of
spins, flavors, and colors, and zero masses, keeps $\Omega_{0}$ invariant
under global U(1) gauge transformations and rotations in color and flavor
space; under global $U(1)$, and special unitary color and flavor rotations of
the field operators, $\psi \to e^{i\varphi}U_cU_f\psi$, the $\phi_\pm$
transform as $(\phi_{\pm})_{abij}\to e^{-2i\varphi}(\phi_{\pm})_{cdlm}
(U_c^\dagger)_{ca}(U_c^\dagger)_{db}(U_f^\dagger)_{li}(U_f^\dagger)_{mj}$.  
Taking into account the condition (\ref{pauli1}) imposed by the Pauli 
principle, we obtain, up to ${\cal O}(\Delta^4)$,
\begin{eqnarray}
  \Omega_{0}&=& \alpha^{+}{\rm Tr}(\phi_{+}^{\dagger}\phi_{+})_{F}
  +\beta^{+}_{1}[{\rm Tr}(\phi_{+}^{\dagger}\phi_{+})_{F}]^{2}
  +\beta^{+}_{2}{\rm Tr}[(\phi_{+}^{\dagger}\phi_{+})^{2}]_{F}
 \nonumber \\
  & & +\alpha^{-}{\rm
  Tr}(\phi_{-}^{\dagger}\phi_{-})_{F} +\beta^{-}_{1}[{\rm
  Tr}(\phi_{-}^{\dagger}\phi_{-})_{F}]^{2} +\beta^{-}_{2}{\rm
  Tr}[(\phi_{-}^{\dagger}\phi_{-})^{2}]_{F}\ .
 \label{omega0}
\end{eqnarray}
Due to the relation, $\Lambda^{\pm}\Lambda^{\mp}=0$, the contributions of
the quark-quark condensates separate from those of the antiquark-antiquark
condensates.  The subscript ``$F$" denotes the pairing gap evaluated for the
quark or antiquark quasiparticle momenta $|{\bf k}|$ equal to the quark Fermi
momentum, $k_F$.

    In general, the energies of antiquark excitations, $\sim \left((|{\bf
k}|+k_F)^2+\phi_-^\dagger\phi_-\right)^{1/2}$, are much larger than the
energies of quark excitations, $\sim \left((|{\bf k}|
-k_F)^2+\phi_+^\dagger\phi_+\right)^{1/2}$.  Thus in the temperature region
near $T_{c}$ where $\mu/3$ is large compared with the magnitude of the energy
gap, the amplitudes for virtual excitations of antiquark quasiparticles with
$|{\bf k}|$ near $k_F$ are smaller than those for excitations of quark
quasiparticles by a factor $\sim{\cal O}(|\phi_{+}(T\sim T_{c})|/\mu)$; as a
consequence, $\alpha^{+}$ and $\beta^{+}_{i}$ dominate over $\alpha^{-}$ and
$\beta^{-}_{i}$, and we set $\alpha^{-}=\beta^{-}_{i}=0$.

    The next step is to express the correction term $\Omega_{\rm CN}$ in terms
of $\phi_{+}$ and ${\tilde\mu}_{ab}$.  We need retain contributions up to
second order in $\tilde\mu_{ab}$.  Since $n_a =-\partial\Omega_{\rm
SF}/\partial \mu_a$ and $n_{ab}=-\partial\Omega_{\rm
SF}/\partial{\tilde\mu}_{ba}$ for $a\neq b$, the conditions for color
neutrality, Eqs.\ (\ref{colsindia}) and (\ref{colsinnondia}), can be written
as
\begin{equation}
 \frac{\partial\Delta\Omega}{\partial {\tilde\mu}_{RR}}=
 \frac{\partial\Delta\Omega}{\partial {\tilde\mu}_{GG}}=
 \frac{\partial\Delta\Omega}{\partial {\tilde\mu}_{BB}}\ ,
  \label{neut}
\end{equation}
and
\begin{equation}
 \frac{\partial\Delta\Omega}{\partial {\tilde\mu}_{ab}}=0
  \label{neutnd}
\end{equation}
for $a\neq b$.

    Up to ${\cal O}({\tilde\mu}_{ab}\phi_{+}^{2})$ and ${\cal O}(
{\tilde\mu}_{ab}^{2})$, the only terms in
$\Omega_{\rm CN}$ that are invariant under global U(1) gauge transformations
and special unitary transformations in flavor space are proportional to ${\rm
Tr}{\tilde{\cal M}}^{2}$, ${\rm Tr}(\phi_{+}^{\dagger}\phi_{+}{\tilde{\cal
M}})_{F}$, and ${\rm Tr}(\phi_{+}\phi_{+}^{\dagger}{\tilde{\cal M}})_{F}$,
where
\begin{equation}
   \tilde{\cal M}_{abij}=\delta_{ij}{\tilde\mu}_{ab}\ ;
  \label{calM}
\end{equation}
in deducing this structure we have used the relations ${\rm
Tr}(\phi_{+}\phi_{+}^{\dagger} {\tilde{\cal M}}^{T})_{F}={\rm
Tr}(\phi_{+}^{\dagger}\phi_{+} {\tilde{\cal M}})_{F}$ and ${\rm
Tr}(\phi_{+}^{\dagger}\phi_{+} {\tilde{\cal M}}^{T})_{F}={\rm
Tr}(\phi_{+}\phi_{+}^{\dagger} {\tilde{\cal M}})_{F}$, derived from
$(\phi_{+}^{T})_F=(\phi_{+})_F$ (see Eq.\ (\ref{pauli1})).  The term
proportional to ${\rm Tr}(\phi_{+}\phi_{+}^{\dagger}{\tilde{\cal M}})_{F}$,
which does not occur in the weak coupling limit (see Eq.~(\ref{loop01})),
always has a zero coefficient, even in the strong coupling regime.  The reason
is that this term, of second order in the gap, is not affected by the
dependence of the pairing interaction on the pairing gap, which can only
induce terms of at least fourth order in the gap.  We thus write
\begin{equation}
  \Omega_{\rm CN}=\sigma{\rm Tr}{\tilde{\cal M}}^{2}
  +\chi{\rm Tr}(\phi_{+}^{\dagger}\phi_{+}{\tilde{\cal M}})_{F}\ .
\label{omegacn}
\end{equation}
The first term on the right side of Eq.\ (\ref{omegacn}) comes from the
change in the normal-fluid free-energy for non-zero $\tilde{\cal M}$, while
the second is the correction to the ${\cal O}(\phi_{+}^{2})$ term in
$\Omega_{0}$.  Because of color neutrality, $\sigma$ and $\chi$ appear, as we
shall see, only in the dimensionless combination $\chi^{2}/\sigma$ in the
energy gap and the condensation energy.

    In the remainder of this section, we consider condensates that are
antisymmetric in color and flavor.   In these condensates, Cooper pairing
is characterized by the products between the color-antitriplet states
[$(|RG\rangle-|GR\rangle)/\sqrt{2}$, $(|GB\rangle-|BG\rangle)/\sqrt{2}$,
$(|BR\rangle-|RB\rangle)/\sqrt{2}$] and the flavor-antitriplet states
[$(|ud\rangle-|du\rangle)/\sqrt{2}$, $(|ds\rangle-|sd\rangle)/\sqrt{2}$,
$(|su\rangle-|us\rangle)/\sqrt{2}$].

    For total angular momentum $J=0$, the Pauli principle constraint
(\ref{pauli1}) requires that antisymmetry of pairing in color space be
accompanied by antisymmetry of pairing in flavor space.  Such pairing states
can occur in the weak coupling limit, because one-gluon exchange in the color
antitriplet channel is attractive \cite{BL}.  The gap is characterized by the
ansatz,
\begin{equation}
  (\phi_{+})_{abij}=\epsilon_{ijl}\epsilon_{abc}A_{lc}\ ,
  \label{icat}
\end{equation}
where $A\equiv({\bf d}_R,{\bf d}_G,{\bf d}_B)$ is a general complex matrix.
$A_{ia}$, $=({\bf d}_a)_{i}$, denotes the gap for pairing between quarks of
colors $b$ and $c$ with $a \ne b \ne c$ and of flavors $j$ and $l$ with
$i \ne j \ne l$.

    Substituting Eq.\ (\ref{icat}) into Eqs.\ (\ref{omega0}) and
(\ref{omegacn}), we derive the Ginzburg-Landau form of $\Omega_{0}$ and
$\Omega_{\rm CN}$ in terms of $A$ or ${\bf d}_{a}$:
\begin{eqnarray}
   \Omega_{0} &=&
      {\bar\alpha}{\rm Tr}(A^{\dagger}A)_{F}
      +\beta_{1}[{\rm Tr}(A^{\dagger}A)_{F}]^{2}
      +\beta_{2}{\rm Tr}[(A^{\dagger}A)^{2}]_{F}
   \nonumber \\ &=&
      {\bar\alpha}\lambda
      +(\beta_{1}+\beta_{2}\Upsilon)\lambda^{2}\ ,
  \label{omega0d}
\end{eqnarray}
with
\begin{equation}
   {\bar\alpha}\equiv4\alpha^{+}\ , ~~
   \beta_{1}\equiv16\beta_{1}^{+}+2\beta_{2}^{+}\ , ~~
   \beta_{2}\equiv2\beta_{2}^{+}\ ,
\end{equation}
\begin{equation}
   \lambda\equiv\sum_{a}|{\bf d}_{a}|_{F}^{2}\ , ~~
   \Upsilon\equiv\frac{1}{\lambda^{2}}
    \sum_{ab}|{\bf d}_{a}^{*}\cdot{\bf d}_{b}|_{F}^{2}\ ,
  \label{zetalambda}
\end{equation}
and
\begin{equation}
  \Omega_{\rm CN}=3\sigma\sum_{ab}|{\tilde\mu}_{ab}|^{2}
        -2\chi\sum_{ab}({\bf d}_{a}^{*}\cdot{\bf d}_{b})_{F}
         {\tilde \mu}_{ab}\ .
  \label{omegacnd}
\end{equation}
One can readily show that under a global $U(1)$, color, and flavor rotation of
the field operators, $\psi \to e^{i\varphi}U_cU_f\psi$, $A$ transforms as 
$A_{ia}\to e^{-2i\varphi}(U_c)_{ab}(U_f)_{ij} A_{jb}$.  The second order 
and fourth order terms in $A$, included in $\Omega_{0}$, Eq.\ (\ref{omega0d}),
are the only invariants under these transformations.  In $\Omega_{\rm CN}$,
Eq.\ (\ref{omegacnd}), the term of linear order in ${\tilde \mu}_{ab}$ is not
affected by global U(1) gauge transformations, ${\bf d}_{a}\rightarrow
e^{-2i\varphi}{\bf d}_{a}$, and flavor rotations, ${\bf d}_{a}\rightarrow 
U_f {\bf d}_{a}$.

    Note that $\Upsilon$, which is dimensionless, ranges from 1/3 to 1.
For $\Upsilon=1/3$, the configurations for ${\bf d}_{a}$ are determined by
\begin{equation}
 {\bf d}_{R}^{*}\cdot{\bf d}_{G}={\bf d}_{G}^{*}\cdot{\bf d}_{B}
     ={\bf d}_{B}^{*}\cdot{\bf d}_{R}=0\ , ~~
 |{\bf d}_{R}|^{2}=|{\bf d}_{G}|^{2}=|{\bf d}_{B}|^{2}\ ;
 \label{opcfl}
\end{equation}
for $\Upsilon=1$, the vectors ${\bf d}_{R}$, ${\bf d}_{G}$, and ${\bf d}_{B}$
are all parallel:
\begin{equation}
  {\bf d}_{R}\parallel{\bf d}_{G}\parallel{\bf d}_{B}\ .
 \label{opis}
\end{equation}

     In the weak coupling limit, which we calculate in Sec.\ \ref{sec:GL},
the coefficients ${\bar\alpha}$, $\beta_{1}$, $\beta_{2}$, $\sigma$, and
$\chi$ reduce to
\begin{equation}
  {\bar\alpha}=4N(\mu/3)\ln\left(\frac{T}{T_{c}}\right)\ ,
 \label{alpha}
\end{equation}
\begin{equation}
  \beta_{1}=\frac{7\zeta(3)}{8(\pi T_{c})^{2}} N(\mu/3)\ ,
 \label{beta1}
\end{equation}
\begin{equation}
  \beta_{2}=\frac{7\zeta(3)}{8(\pi T_{c})^{2}} N(\mu/3)\ ,
 \label{beta2}
\end{equation}
\begin{equation}
   \sigma=-N(\mu/3)\ ,
  \label{sigma}
\end{equation}
\begin{equation}
   \chi=\frac{3}{\mu}\ln\left(\frac{3T_{c}}{\mu}\right)
  N(\mu/3)\ ,
  \label{chi}
\end{equation}
where
\begin{equation}
  N(\mu/3) =\frac{1}{2\pi^{2}}
   \left(\frac{\mu}{3}\right)^{2}
\end{equation}
is the ideal gas density of states at the Fermi surface, and the zeta function
$\zeta(3)=1.2020\ldots$.

    Two effects beyond weak coupling must in general be taken into account.
The first is the modification of the pairing interaction due to the
pairing gap, which modifies the coefficients $\beta_{1}$ and $\beta_{2}$ of
the fourth order terms in $A$.  The second is radiative corrections by
the normal medium \cite{FM} -- i.e., quark self-energy, gluon polarization,
and quark-quark-gluon and three-gluon vertex corrections -- which modify the
coefficients Eqs.\ (\ref{alpha})-(\ref{chi}), mainly through their dependence
on $N(\mu/3)$ and $T_{c}$.

    We proceed to minimize the Ginzburg-Landau free energy, Eqs.\
(\ref{omega0d}) and (\ref{omegacnd}), with respect to the $({\bf d}_{a})_{F}$.
To elucidate effects of the color neutrality, it is instructive to start with
the optimal expressions for the energy gap and the condensation energy in the
case in which the ${\tilde\mu}_{ab}$ and hence ${\tilde{\cal M}}$ and
$\Omega_{\rm CN}$ vanish.  From $\Omega_0$, written up to second order in
$\lambda$, we find that thermodynamic stability requires
${\bar\beta}\equiv\beta_{1}+ \beta_{2}\Upsilon>0$ for $1/3\leq\Upsilon\leq1$.
It is straightforward to show that in this stable region, only two phases
occur:  $\Upsilon=1/3$ for $\beta_{2}>0$ and $\Upsilon=1$ for $\beta_{2}<0$.
We discuss the physics of the region where ${\bar\beta}<0$ below.

    The order parameters in the $\Upsilon=1$ phase satisfy condition
(\ref{opis}).  All such order parameters lead to states degenerate in energy.
Note that in this order-parameter set, any state is identical to an isoscalar,
color-antitiplet state characterized by $({\bf d}_{a})_{i}\propto \delta_{is}$
($s$, the strange flavor), or equivalently,
\begin{equation}
  (\phi_{+})_{abij}=\epsilon_{abc}\epsilon_{ijs}({\bf d}_{c})_{s}\ ,
  \label{icat1}
\end{equation}
to within a constant phase factor and a special unitary
transformation in flavor space.  In the corresponding condensate, the $u$ and
$d$ quarks are paired in an isosinglet state.  We shall refer to the
$\Upsilon=1$ phase as the isoscalar phase, even though it contains
order parameters having the other orientations in complex flavor space than
the $s$ direction.  The magnitude of the gap in the isoscalar phase is,
\begin{equation}
  \lambda=-\frac{{\bar\alpha}}{2(\beta_{1}+\beta_{2})}\ ,
  \label{lambdaIS}
\end{equation}
with condensation energy,
\begin{equation}
\Delta\Omega=-\frac{{\bar\alpha}^{2}}{4(\beta_{1}+\beta_{2})}\ .
 \label{EcondIS}
\end{equation}

    The order parameters in the $\Upsilon=1/3$ phase fulfill condition
(\ref{opcfl}), and lead to degenerate states.  This phase is the color-flavor
locked phase \cite{ARW}; the condensate in this phase is characterized by its
symmetry under simultaneous exchange of color and flavor.  States belonging to
this phase transform into one another under global U(1) gauge transformations
and special unitary transformations in flavor space.  The simplest among these
states is described by $({\bf d}_{a})_{i}\propto \delta_{ai}$ and $({\bf
d}_{R})_{u}=({\bf d}_{G})_{d}=({\bf d}_{B})_{s} \equiv\kappa_{A}$; the
corresponding gap matrix is given explicitly by
\begin{equation}
  (\phi_{+})_{abij}=\kappa_{A}
  (\delta_{ai}\delta_{bj}-\delta_{aj}\delta_{bi})\ .
 \label{phicfl}
\end{equation}
For the color-flavor locked phase, we obtain the magnitude of the gap,
\begin{equation}
  \lambda=-\frac{{\bar\alpha}}{2(\beta_{1}+\beta_{2}/3)}\ ,
  \label{lambdaCFL}
\end{equation}
and the condensation energy,
\begin{equation}
\Delta\Omega=-\frac{{\bar\alpha}^{2}}{4(\beta_{1}+\beta_{2}/3)}\ .
 \label{EcondCFL}
\end{equation}

    In the original definition \cite{ARW} of the color-flavor locked state,
in addition to the gap, $\kappa_{A}$, in the color and flavor antisymmetric
channel, a gap matrix $\kappa_{S}(\delta_{ai}\delta_{bj}+\delta_{aj}
\delta_{bi})$ arises in the color and flavor symmetric channel.  In the weak
coupling limit, the gap $\kappa_{A}$ can be generated by the attractive
one-gluon exchange interaction in the color-antitriplet channel.  Nonzero
values of $\kappa_{S}$, on the other hand, are not driven by the one-gluon
exchange interaction since it is repulsive in the color symmetric channel;
they are ${\cal O}(g)$ correction to $\kappa_{A}$ that ensures the existence
of the $\kappa_{A}\neq0$ solution to the relevant weak coupling gap equation
(see Ref.\ \cite{tom}).   This implies that the gap $\kappa_{S}$ can be
ignored in the temperature region near the onset of the pairing.

    We now construct, in the $\beta_{1}$-$\beta_{2}$ plane, the
phase diagram exhibiting the more stable pairing state of the isoscalar,
and color-flavor locked phases.  We first address the question of whether or
not $T_{c}$, the temperature at which the pairing instability of the normal
phase occurs, is the same for these phases.  The onset of this instability is
controlled solely by the pairing interactions in the normal phase between
quark quasiparticles with zero total momentum.  The corresponding amplitudes
depend on the color, flavor, and quantum numbers such as the total angular
momentum, chirality, and parity of the quarks involved in the pairing
\cite{BL}, and for the two phases considered here, the quark-quark pairs have
antisymmetric structures in color and flavor, the same chirality, even
parity, and $J=0$ at the onset of the pairing instability.  We thus conclude
that no difference in $T_{c}$ arises between these phases.  The amplitudes in
the normal system driving the instability do not distinguish between the final
possible paired states.

    Figure 1 shows the map of the isoscalar and color-flavor locked phases in
the $\beta_{1}$-$\beta_{2}$ plane, for the case in which ${\tilde{\cal M}}$,
or equivalently, $\Gamma$, is set equal to zero.  The regions subtended by
these phases are restricted by the region in which ${\bar\beta}>0$, for
$1/3\leq\Upsilon\leq1$, is violated.  By comparing the condensation energies
Eqs.\ (\ref{EcondIS}) and (\ref{EcondCFL}), we find that in the weak coupling
limit, described here by Eqs.\ (\ref{alpha})-(\ref{beta2}), the color-flavor
locked state is favored over the isoscalar channel, a result consistent with
the conclusion drawn from the weak coupling analyses \cite{tom,EHHS} at zero
temperature.

\begin{figure}
\begin{center}
\epsfig{file=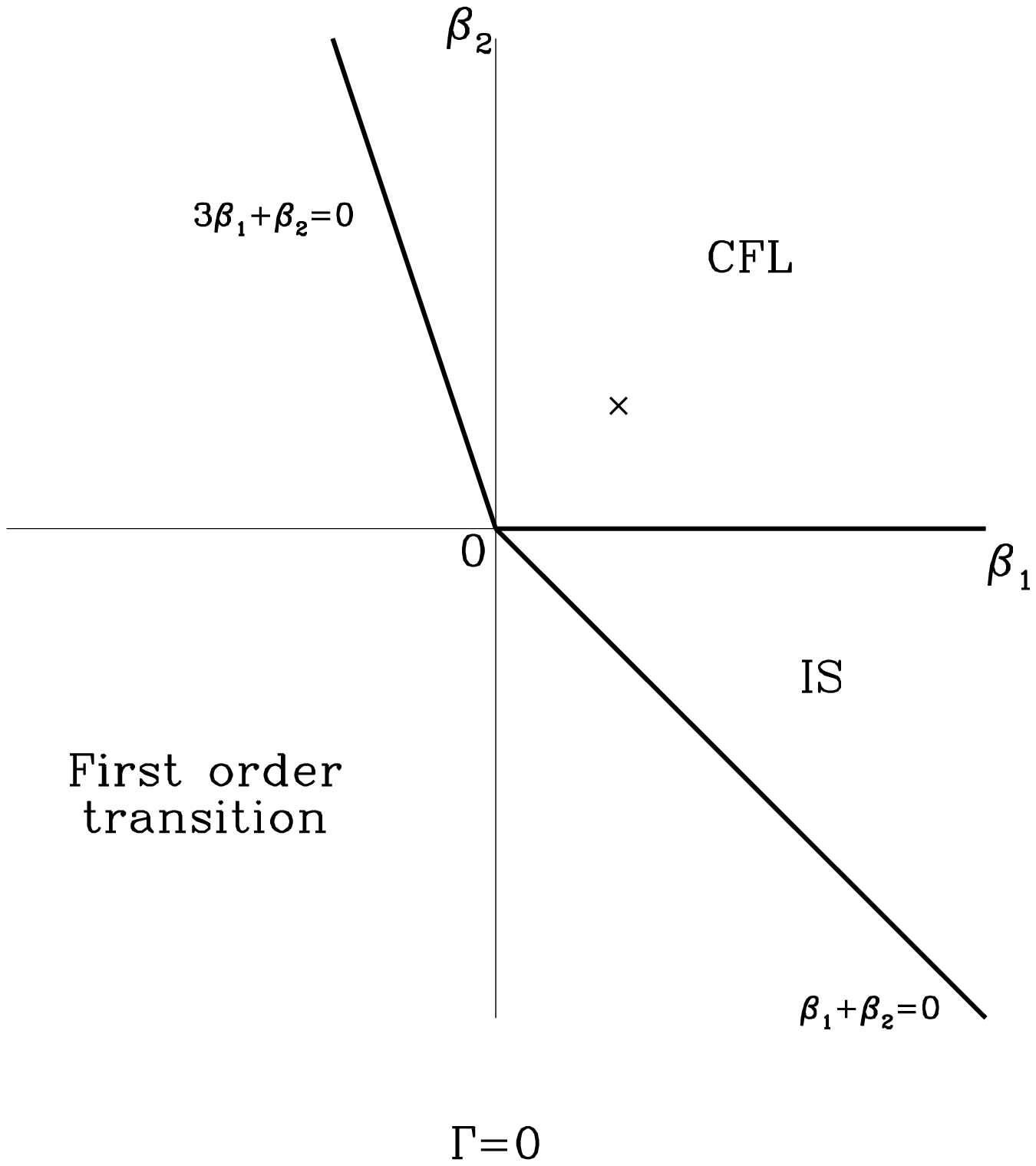, height=8.5cm}
\end{center}
Fig.~1.  Phase diagram in the Ginzburg-Landau regime, showing
regions where the isoscalar (IS) and color-flavor locked (CFL) phases are
favored, when the constraint of color neutrality does not affect the free
energy, $\Gamma=0$.  The $\beta_{1}$, $\beta_{2}$ are the fourth order
coefficients in the Ginzburg-Landau free energy, Eq.~(\ref{omega0d}),
and $\Gamma$ (Eqs.~(\ref{gamma}) and (\ref{omegacnd})) describes
effects of the constraint of color neutrality.  The cross indicates the weak
coupling limit.  In the region of first order transitions, the overall
fourth order coefficient, $\bar\beta$, is not positive definite.
\end{figure}

    The region in which ${\bar\beta}<0$ corresponds to a {\it first order}
phase transition from the normal to the superfluid state at a temperature
greater than $T_c^0$ at which ${\bar\alpha}(T_{c}^{0})$ vanishes, and the
normal state becomes unstable against Cooper pairing.  The situation is
similar to condensed matter systems in which the Ginzburg-Landau free energy
contains a term cubic in the magnitude of the order parameter, and also
similar to the situation in which the chiral phase transition in QCD with 
two-flavor massless quarks changes from second order to first order at a 
tricritical point in the $\mu$ versus $T$ plane \cite{tricritical}.  To see 
this structure, we add to the Ginzburg-Landau expansion up to fourth order in
$A$, Eq.\ ({\ref{omega0d}}), the sum of the sixth order invariants,
\begin{equation}
\Omega_{6}=({\bar\gamma}_{1}+{\bar\gamma}_{2}\Upsilon+{\bar\gamma}_{3}\xi)
           \lambda^{3}\ ,
\end{equation}
where $\lambda$ and $\Upsilon$ are given by Eq.\ (\ref{zetalambda}), and
\begin{equation}
 \xi\equiv\frac{1}{\lambda^{3}}
  \sum_{abc}({\bf d}_{a}^{*}\cdot{\bf d}_{b})_F
  ({\bf d}_{b}^{*}\cdot{\bf d}_{c})_F ({\bf d}_{c}^{*}\cdot{\bf d}_{a})_F\ .
   \label{xi}
\end{equation}
Let us assume that the overall coefficient of $\lambda^{3}$,
${\bar\gamma}_{1} +{\bar\gamma}_{2}\Upsilon +
{\bar\gamma}_{3}\xi\equiv{\bar\gamma}$, is positive.  Then, the local minimum
of $\Omega_{0}+\Omega_{6}$ is reached at a nonzero value of $\lambda$ that can
be calculated as $\lambda_{0}=|{\bar\beta}|
/3{\bar\gamma}+[({\bar\beta}/3{\bar\gamma})^{2}
-{\bar\alpha}/3{\bar\gamma}]^{1/2}$.  The critical temperature $T_{c}$ can be
determined from the condition
$\Omega_{0}(\lambda_{0})+\Omega_{6}(\lambda_{0})=0$, or equivalently,
${\bar\alpha}(T_{c})={\bar\beta}^{2}/4{\bar\gamma}$, where we can ignore the
temperature dependence of ${\bar\beta}$ and ${\bar\gamma}$.  Note that $T_c$
is greater than $T_{c}^{0}$.  We thus find that $\lambda_{0}
=|{\bar\beta}|/2{\bar\gamma}$ at $T=T_{c}$, i.e., the pairing gap is
discontinuous at the transition point.  We remark that this argument is only
applicable to the case in which $T_{c}-T_{c}^{0} \ll T_{c}^{0}$.  In the case
that ${\bar\beta}<0$ and ${\bar\gamma}<0$, one must go to higher order to
determine the critical temperature and the discontinuity of the order
parameter at the transition.

    We turn to the Ginzburg-Landau energy, Eqs.\ (\ref{omega0d}) and
(\ref{omegacnd}), for a color-singlet system.  As we shall see,
the constraint of color neutrality acts to modify the pairing gap without
removing the degeneracy of the order parameters occurring when the differences
in color chemical potentials vanish.
Generally, the deviations of the chemical potential differences
${\tilde\mu}_{ab}$ from zero can be determined from the color neutrality
conditions (\ref{neut}) and (\ref{neutnd}) as
\begin{equation}
   {\tilde\mu}_{ab}=\frac{\chi}{9\sigma}
   [3({\bf d}_{a}\cdot{\bf d}_{b}^{*})_{F}-
   \delta_{ab}\sum_{c}|{\bf d}_{c}|_{F}^{2}]\ .
  \label{mua}
\end{equation}
Substitution of this expression into Eqs.\ (\ref{omega0d}) and
(\ref{omegacnd}) leads to the Ginzburg-Landau form,
\begin{equation}
  \Delta\Omega={\bar\alpha}\lambda
      +[(\beta_{1}-\Gamma)+(\beta_{2}+3\Gamma)\Upsilon]\lambda^{2}\ ,
    \label{delomegad}
\end{equation}
with
\begin{equation}
   \Gamma\equiv-\frac{1}{\sigma}\left(\frac{\chi}{3}\right)^{2}\ .
  \label{gamma}
\end{equation}
The coefficient $\Gamma$ is positive definite as long as the color-singlet
system is thermodynamically stable against color fluctuations, i.e.,
$(\partial^{2}\Delta\Omega/ \partial{\tilde\mu}_{ab}\partial
{\tilde\mu}_{ba})_{({\bf d}_{c})_{F}}=6\sigma<0$.  This condition holds in the
weak coupling limit, Eq.\ (\ref{sigma}); vacuum polarization effects in the
low-density regime, which lead to color antiscreening, or positive color
susceptibility, $-\partial^{2}\Omega/\partial{\tilde\mu}_{ab}\partial
{\tilde\mu}_{ba}>0$, are in the direction to preserve this condition.  We thus
assume $\Gamma>0$ for arbitrary coupling constant $g$.

    Note that Eq.\ (\ref{delomegad}) is identical with Eq.\ (\ref{omega0d})
except that $\beta_{1}$ and $\beta_{2}$ are replaced by $\beta_{1}-\Gamma$ and
$\beta_{2}+3\Gamma$, i.e., inclusion of the color neutrality results in a
renormalization of $\beta_{1}$ and $\beta_{2}$.  This is due to the fact that
the constraint of color neutrality itself is invariant under color rotations 
and thus chemical potential differences between colors yield the fourth order 
terms invariant with respect to rotations in color space.

    We then obtain from Eq.\ (\ref{delomegad}) the two sets of optimal order
parameters, $\Upsilon=1/3$ (the color-flavor locked phase) for
$\beta_{2}+3\Gamma>0$ and $\Upsilon=1$ (the isoscalar phase) for
$\beta_{2}+3\Gamma<0$.  The resultant magnitude of the gap in the isoscalar
phase can be written as
\begin{equation}
  \lambda=-\frac{{\bar\alpha}}{2(\beta_{1}+\beta_{2}+2\Gamma)}\ ,
  \label{lambdaIS1}
\end{equation}
with condensation energy,
\begin{equation}
\Delta\Omega=-\frac{{\bar\alpha}^{2}}{4(\beta_{1}+\beta_{2}+2\Gamma)}\ .
 \label{EcondIS1}
\end{equation}
We observe from the energy gap and condensation energy, Eqs.\
(\ref{lambdaIS1}) and (\ref{EcondIS1}), that $\lambda$ and
$|\Delta\Omega|$ are suppressed at fixed $\beta_{1}$ and $\beta_{2}$ by the
parameter $\Gamma$ characterizing the color neutrality.  This suppression,
together with ${\tilde{\cal M}}\neq0$ as can be found from substitution of
condition (\ref{opis}) into Eq.\ (\ref{mua}),
comes from the fact that the condensate is intrinsically anisotropic in color
space, as we shall see in Sec.\ \ref{subsec:GE2SC}.  We remark in passing that
in the weak coupling limit, where only contributions of leading order in
$3T_{c}/\mu$ remain, $\Gamma$ is dominated by $\beta_{1}$ and $\beta_{2}$
and hence does not significantly affect the gap.  In the color-flavor locked
phase, on the other hand, the gap size and the condensation energy of the
color-flavor locked phase are still given by Eqs.\ (\ref{lambdaCFL}) and
(\ref{EcondCFL}).  For this phase, we can find ${\tilde{\cal M}}=0$ from
substitution of condition (\ref{opcfl}) into Eq.\ (\ref{mua}).  This
vanishing reflects the fact that the condensate in the color-flavor locking is
isotropic in color space, as suggested by condition (\ref{opcfl}).

    We conclude this section by examining the influence of the requirement of
color neutrality on the phase diagram in the $\beta_{1}$-$\beta_{2}$ plane.
As we see in Fig.\ 2, the phase diagram for $\Gamma>0$, this requirement
shifts the isoscalar phase by $\Gamma$ in the $\beta_{1}$ direction and
$-3\Gamma$ in the $\beta_{2}$ direction.  With increasing $\Gamma$, this
phase, in which the order parameters are anisotropic in color space, moves
farther away from the weak coupling point, $(\beta_{1},\beta_{2})$ given by
Eqs.\ (\ref{beta1}) and (\ref{beta2}).  With increasing $\Gamma$, the
color-flavor locked phase begins to cover the region occupied by the isoscalar
phase.

    To answer the question of which phase is more stable at low densities
will require understanding in detail how strong coupling effects, which
develop as the baryon chemical potential $\mu$ is lowered, modify $\beta_{1}$,
$\beta_{2}$, and $\Gamma$ with increasing $g$.  These effects are too
uncertain for us to be able to predict in general the more stable state near
$T_{c}$.

\begin{figure}
\begin{center}
\epsfig{file=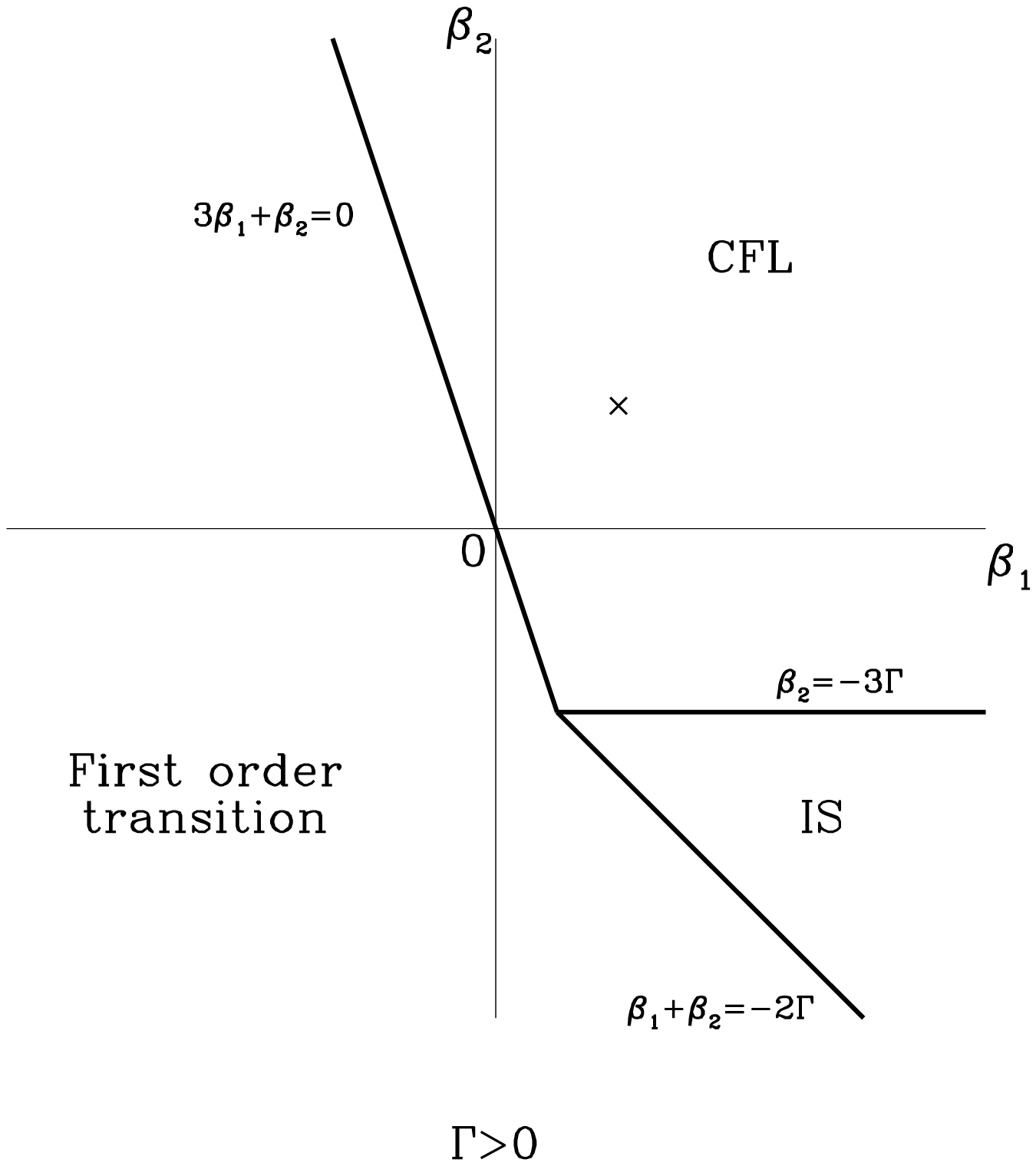, height=8.5cm}
\end{center}
    Fig.~2.  Same as Fig.\ 1 with the constraint of color neutrality,
described by $\Gamma>0$.
\end{figure}

\section{Gap equation}
\label{sec:GE}

    We turn now to deriving the high-density asymptotic form of the
thermodynamic potential difference $\Delta\Omega$ near $T_{c}$, as we used in
the previous section.  The gap equation in the weak coupling limit provides
information on the overall temperature dependence of the gap and the
quasiparticle structures, in contrast to the Ginzburg-Landau approach
developed in the previous section, which concentrates on limited temperature
region near $T_c$ and coarse-grained features of the superfluid.  For the
moment, we consider an arbitrary color neutral condensate of the form
(\ref{delta1}).  For the two optimal pairing states considered in Sec.\
\ref{sec:GGL}, we obtain the behavior of the energy gap from the
finite-temperature gap equation, with the constraint of color
neutrality.  In Sec.\ \ref{sec:GL}, we derive the coefficients given by
Eqs.\ (\ref{alpha})-(\ref{chi}) from the structure of the energy gap.

    In obtaining the gap equation in the weak coupling limit, it is convenient
to first introduce the notation for the quark field, $(\psi_{ai},
\psi^{C}_{ai})$.  We then write a self-consistent Schwinger-Dyson equation to
determine the proper self energy $\Sigma(k)$ up to ${\cal O}(g^{2})$
\cite{BL}:
\begin{eqnarray}
  \Sigma(k)
  &=&-g^{2}T\int\frac{d^{3}q}{(2\pi)^{3}}\sum_{n~ {\rm odd}}
   D_{\mu\nu}^{\alpha\beta}(k-q) \hspace{150pt}
  \nonumber \\
 &&\times   \pmatrix{
   \gamma^{\mu}\lambda^{\alpha}/2 & 0 \cr
   0 & -\gamma^{\mu}(\lambda^{\alpha})^{T}/2 \cr
    }
   G(q)
    \pmatrix{
   \gamma^{\nu}\lambda^{\beta}/2 & 0 \cr
   0 & -\gamma^{\nu}(\lambda^{\beta})^{T}/2 \cr
    }\ ,
  \label{self}
\end{eqnarray}
where the summation is over Matsubara frequencies, $q_{0}=in\pi T$, with
$n$ odd for fermions,
\begin{eqnarray}
  G(k) &\equiv&
  \pmatrix{
  G^{(11)}(k)& G^{(12)}(k) \cr
  G^{(21)}(k)& G^{(22)}(k) \cr
    }
  =  \pmatrix{
  \gamma k+\gamma^{0}{\cal M}  & {\tilde \Delta}(k) \cr
  \Delta(k) & \gamma k-\gamma^{0}{\cal M}^{T} \cr
      }^{-1}
  \label{G}
\end{eqnarray}
is the quark propagator with normal state Hartree-Fock contributions
ignored, ${\tilde \Delta}=\gamma^{0}\Delta^{\dagger}\gamma^{0}$ and ${\cal
M}_{abij}=\delta_{ij}(\delta_{ab}\mu/3+{\tilde\mu}_{ab})$, and
$D_{\mu\nu}^{\alpha\beta}$ is the gluon propagator, specified below.

    The self-energy and quark propagator are related by
\begin{equation}
   G^{-1}(k)=G_{0}^{-1}(k) - \Sigma(k)\ ,
 \label{GSig}
\end{equation}
where
\begin{equation}
  G_{0}(k)=
 \pmatrix{
 \gamma k+\gamma^{0}{\cal M}  & 0 \cr
  0 & \gamma k-\gamma^{0}{\cal M}^{T} \cr
    }^{-1}
 \label{G0}
\end{equation}
is the noninteracting quark propagator.

    The off-diagonal component of Eq.\ (\ref{self}) yields the gap equation
in the weak coupling limit,
\begin{equation}
  \Delta(k)=-g^{2}T\int\frac{d^{3}q}{(2\pi)^{3}}\sum_{n~ {\rm odd}}
    D_{\mu\nu}^{\alpha\beta}(k-q)
   \gamma^{\mu}\frac{(\lambda^{\alpha})^{T}}{2}
     G^{(21)}(q)
   \gamma^{\nu}\frac{\lambda^{\beta}}{2}\ ,
   \label{wcgapeq}
\end{equation}
where
\begin{equation}
  G^{(21)}(q)=(\gamma q-\gamma^{0}{\cal M}^{T})^{-1}\Delta(q)
  \left[{\tilde \Delta}(q)(\gamma q-\gamma^{0}{\cal M}^{T})^{-1}\Delta(q)
   -(\gamma q+\gamma^{0}{\cal M})\right]^{-1}
  \label{G21}
\end{equation}
is the 21-component of $G(q)$.  The summation over $n$ on the right side
of Eq.\ (\ref{wcgapeq}) leads to the self-consistent gap equation,
\begin{eqnarray}
  \Delta(\omega_{\pm}({\bf k}),{\bf k})&=&
  -\frac{1}{4}g^{2}\sum_{\eta=\pm}\eta\int\frac{d^{3}q}{(2\pi)^{3}}
   D_{\mu\nu}^{\alpha\beta}(\varepsilon_{\pm}({\bf k})
    -\eta\varepsilon_{\eta}({\bf q}),{\bf k-q})
   \nonumber \\ & &
  \times\gamma^{\mu}\frac{(\lambda^{\alpha})^{T}}{2}
   \left[{\hat \Delta}(\omega_{\eta}({\bf q}),{\bf q})
   \varepsilon^{-1}_{\eta}({\bf q})
   \tanh\left(\displaystyle{\frac{\omega_{\eta}({\bf q})}
   {2T}}\right)\right]\gamma^{\nu}\frac{\lambda^{\beta}}{2}\ ,
   \label{bcsgap}
\end{eqnarray}
where the (on-shell) gap matrix ${\hat\Delta}(k)$ is given by,
\begin{equation}
  {\hat\Delta}(k)=
    \Lambda^{-}({\bf {\hat k}})\gamma^{0}\Delta(k)
    \Lambda^{+}({\bf {\hat k}})\gamma^{0}\ ,  ~~~
\end{equation}
with the Hermitean frequency matrix
\begin{equation}
  \omega_{\pm}({\bf k})=\pm\varepsilon_{\pm}({\bf k})
    -i{\rm Im}{\cal M}\ ,
\end{equation}
and
\begin{eqnarray}
  (\varepsilon_{\eta}({\bf k}))_{abij} &=&
    \left(\delta_{ij}\sum_{c}(\delta_{ac}
    (|{\bf k}|-\mu/3)-{\rm Re}{\tilde\mu}_{ac})(\delta_{cb}
    (|{\bf k}|-\mu/3)-{\rm Re}{\tilde\mu}_{cb})
    \right.
    \nonumber \\ & &
    \left.
     +\frac{1}{2}\sum_{cl\mu\nu}
    {\tilde \Delta}_{acil}^{\mu\nu}(\omega_{\eta}({\bf k}),{\bf k})
    {\hat \Delta}_{cblj}^{\nu\mu}(\omega_{\eta}({\bf k}),{\bf k})
    \right)^{1/2}
\end{eqnarray}
the energy matrix for quark quasiparticles.  Here, we have used
${\cal M}{\cal M}^{T}=({\rm Re}{\cal M})^{2}
+{\cal O}({\tilde{\cal M}}^{2})$ and made the approximation in
$G^{(21)}(q)$, Eq.\ (\ref{G21}), that $(\gamma q-\gamma^{0}{\cal
M}^{T})^{-1}\Delta(q) (\gamma q-\gamma^{0}{\cal M}^{T}) \approx{\tilde
\Delta}^{\dagger}(q)$.  This approximation is sufficient to describe the gap
matrix up to first order in ${\tilde {\cal M}}$, since $||{\bf q}|-\mu/3|
|{\tilde\mu}_{ab}|$ is much smaller than the square of the gap magnitude in
the momentum region where the gap is appreciable (as we shall see in Sec.\
\ref{subsec:GE2SC}).  We have also disregarded the contributions of antiquark
quasiparticles to the integral over ${\bf q}$ in Eq.\ (\ref{bcsgap}); in the
weak coupling regime, these contributions are suppressed by one power of
$\Delta/\mu$ compared with the quark quasiparticle contributions since the
integral is dominated by the region immediately close to $|{\bf q}|=\mu/3$.
Using the definition (\ref{phi}), we obtain
\begin{equation}
 {\hat\Delta}(k)=-\gamma^{5}\phi_{+}(k)\Lambda^{-}({\bf {\hat k}})\ ,
\end{equation}
and the quasiparticle energies
\begin{equation}
 \varepsilon_{\eta}({\bf k})=[(|{\bf k}|-{\rm Re}{\cal M})^{2}+
 \phi_{+}^{\dagger}(\omega_{\eta}({\bf k}),{\bf k})\phi_{+}
 (\omega_{\eta}({\bf k}),{\bf k})]^{1/2}\ .
  \label{qpenergy}
\end{equation}
Equation (\ref{bcsgap}) can now be rewritten in terms of $\phi_{+}$ as
\begin{eqnarray}
  \phi_{+}(\omega_{\pm}({\bf k}),{\bf k}) &=&
  -\frac{g^{2}}{8}\sum_{\eta=\pm}\eta\int\frac{d^{3}q}{(2\pi)^{3}}
  D_{\mu\nu}^{\alpha\beta}
  (\varepsilon_{\pm}({\bf k})-\eta\varepsilon_{\eta}({\bf q}),{\bf k-q})
  {\rm Tr}[\gamma^{\mu}\Lambda^{-}({\bf {\hat q}})
  \gamma^{\nu}\Lambda^{+}({\bf {\hat k}})]
 \nonumber \\ & &
 \times\frac{1}{4}(\lambda^{\alpha})^{T}
  \phi_{+}(\omega_{\eta}({\bf q}),{\bf q})
  \varepsilon_{\eta}^{-1}({\bf q})
  \tanh\left(\displaystyle{\frac{\omega_{\eta}({\bf q})}{2T}}\right)
   \lambda^{\beta}\ .
  \label{phigap}
\end{eqnarray}

    The gap equation (\ref{phigap}) contains the color chemical potential
differences ${\tilde\mu}_{ab}$, which are determined by the color neutrality
conditions, Eqs.\ (\ref{colsindia}) and (\ref{colsinnondia}), written in terms
of $n_{ab}$ with $n_{a}\equiv n_{aa}$; these $n_{ab}$, characterizing the
color charge densities, are given by
\begin{equation}
  n_{ab}=T\sum_{i}\int\frac{d^{3}q}{(2\pi)^{3}}
  \sum_{n~ {\rm odd}}{\rm Tr}[\gamma^{0}G^{(11)}_{abii}(q)]\ ,
  \label{na}
\end{equation}
where
\begin{equation}
   G^{(11)}(q)= \left[(\gamma q+\gamma^{0}{\cal M})
   -{\tilde \Delta}(q)(\gamma q-\gamma^{0}{\cal M}^{T})^{-1}\Delta(q)
    \right]^{-1}
 \label{G110}
\end{equation}
is the 11-component of the propagator $G(q)$.  In terms of $\phi_{\pm}$,
\begin{eqnarray}
  G^{(11)}(q)&=&(\gamma q-\gamma^{0}{\cal M}^{T})
  \left\{
  [(q_{0}+i{\rm Im}{\cal M})^{2}
  -(|{\bf q}|+{\rm Re}{\cal M})^{2}
  -\phi_{-}^{\dagger}(q)\phi_{-}(q)]^{-1}
  \Lambda^{+}({\bf {\hat q}})  \right.
  \nonumber \\ & &
  \left.  +
  [(q_{0}+i{\rm Im}{\cal M})^{2}
  -(|{\bf q}|-{\rm Re}{\cal M})^{2}
  -\phi_{+}^{\dagger}(q)\phi_{+}(q)]^{-1}\Lambda^{-}({\bf {\hat q}})
  \right\}\ ,
  \label{G11}
\end{eqnarray}
where we have again used the approximation $(\gamma q-\gamma^{0}{\cal
M}^{T})^{-1} \Delta(q) (\gamma q-\gamma^{0}{\cal M}^{T})\approx{\tilde
\Delta}^{\dagger}(q)$.  As in the derivation of the gap equation
(\ref{bcsgap}), this approximation is sufficient to determine $n_{ab}$ up to
first order in ${\tilde {\cal M}}$.  Substituting Eq.\ (\ref{G11}) into Eq.\
(\ref{na}), we find
\begin{eqnarray}
  n_{ab}&=&T\sum_{i}\int\frac{d^{3}q}{(2\pi)^{3}}
  \sum_{n~ {\rm odd}}
  2\left\{(q_{0}-{\cal M}^{T}-|{\bf q}|)
  [(q_{0}+i{\rm Im}{\cal M})^{2}
  -(|{\bf q}|+{\rm Re}{\cal M})^{2}
  -\phi_{-}^{\dagger}(q)\phi_{-}(q)]^{-1}
  \right.
  \nonumber \\ & &
  \left.
  +(q_{0}-{\cal M}^{T}+|{\bf q}|)
  [(q_{0}+i{\rm Im}{\cal M})^{2}
  -(|{\bf q}|-{\rm Re}{\cal M})^{2}
     -\phi_{+}^{\dagger}(q)\phi_{+}(q)]^{-1}
   \right\}_{abii}\ .
   \label{na1}
\end{eqnarray}
The Matsubara frequency summation in this equation yields the usual BCS
expression,
\begin{eqnarray}
  n_{ab}&=&\sum_{i}\int\frac{d^{3}q}{(2\pi)^{3}}
   \{1-u_{+}^2({\bf q}) [1-2f(\omega_{+}({\bf q}))]
   + v_{-}^2({\bf q}) [1-2f(-\omega_{-}({\bf q}))]
  \nonumber \\ & &
   +{\cal O}(\phi_{-}^{\dagger}\phi_{-}/\mu^{2})\}_{abii}\ ,
   \label{na2}
\end{eqnarray}
where
\begin{equation}
   f(\varepsilon)=\frac{1}{e^{\varepsilon/T}+1}
\end{equation}
is the distribution function for quark quasiparticles, and the BCS coherence
factors are given by
\begin{equation}
    u_{\eta}^2({\bf q}) = \frac{1}{2}[\varepsilon_{\eta}({\bf q})
     -{\rm Re}{\cal M}+|{\bf q}|]
     \varepsilon_{\eta}^{-1}({\bf q})\ ,   ~~~
    v_{\eta}^2({\bf q}) = \frac{1}{2}[\varepsilon_{\eta}({\bf q})
     +{\rm Re}{\cal M}-|{\bf q}|]
     \varepsilon_{\eta}^{-1}({\bf q})\ ,
\end{equation}
with $u_{\eta}^2({\bf q}) + v_{\eta}^2({\bf q}) = 1$.  In obtaining
${\tilde{\cal M}}$ up to leading order in ${\phi_{+}}$, Eq.\ (\ref{na2}) can
be simplified as
\begin{eqnarray}
  n_{ab} &=&2\sum_{i}\int\frac{d^{3}q}{(2\pi)^{3}}
   \{f(|{\bf q}|-{\cal M})-f(|{\bf q}|-\mu/3)
  \nonumber \\ & &
   +[u_{+}^2({\bf q})f(\varepsilon_{+}({\bf q}))
   +v_{+}^2({\bf q})(1-f(\varepsilon_{+}({\bf q})))]_{{\tilde{\cal M}}=0}
   \}_{abii}\ ,
   \label{na3}
\end{eqnarray}
where we have used the fact that $ \varepsilon_{+}({\bf q})|_{{\rm
Im}{\cal M}=0} =\varepsilon_{-}({\bf q})|_{{\rm Im}{\cal M}=0}
\equiv\varepsilon({\bf q})$ and thus $\phi_{+}(\varepsilon({\bf q}),{\bf q})=
\phi_{+}(-\varepsilon({\bf q}),{\bf q})$.

    In specifying the gluon propagator $D^{\alpha\beta}(p\equiv k-q)$, it is
essential to take into account the long-range nature of the color magnetic
interactions \cite{BMPR,son,PR3,SW}.  Once the effects of a normal medium are
included in the random-phase approximation (RPA), these interactions are
dynamically screened by the Landau damping of the virtual gluons involved
\cite{BMPR,art}.  Since the dominant contributions from $D(p)$ are peaked
around $p=0$, we focus our attention on $|p_{0}|\ll\mu/3$ and $|{\bf p}|\ll
\mu/3$.  The gluon propagator in the Landau gauge can be written
within the RPA as
\begin{equation}
  D_{\mu\nu}^{\alpha\beta}(p)=-\frac{\delta_{\alpha\beta}P^{T}_{\mu\nu}}
  {-p^{2}+m_{D}^{2}\chi_{T}(p_{0}/|{\bf p}|)}
  -\frac{\delta_{\alpha\beta}P^{L}_{\mu\nu}}
  {-p^{2}+m_{D}^{2}\chi_{L}(p_{0}/|{\bf p}|)}\ ,
  \label{glueprop}
\end{equation}
with the dimensionless transverse and longitudinal polarizations,
\begin{equation}
  \chi_{T}(x)=\frac{x^{2}}{2}+\frac{x(1-x^{2})}{4}
  \ln\left(\frac{x+1}{x-1}\right)\ ,
\end{equation}
\begin{equation}
  \chi_{L}(x)=(1-x^{2})
  \left[1-\frac{x}{2}\ln\left(\frac{x+1}{x-1}\right)\right]\ ,
\end{equation}
and transverse and longitudinal projection operators,
\begin{equation}
  P^{T}_{ij}=\delta_{ij}-\frac{p_{i}p_{j}} {|{\bf p}|^{2}}\ , ~~
  P^{T}_{00}=P^{T}_{0i}=P^{T}_{i0}=0\ ,
\end{equation}
\begin{equation}
    P^{L}_{\mu\nu}=\frac{p_{\mu}p_{\nu}}{p^{2}}-g_{\mu\nu}-P^{T}_{\mu\nu}\ ,
\end{equation}
where the Debye mass is given by
\begin{equation}
   m_{D}=\left(\frac{g^{2}}{6\pi^{2}}{\rm Tr}{\cal M}^{2}
     +\left(3+\frac{N_{f}}{2}\right)\frac{g^{2}T^{2}}{3}\right)^{1/2}\ .
 \label{debye}
\end{equation}

    In Eq.\ (\ref{glueprop}), the term proportional to $P^{T}$ describes the
transverse or color magnetic sector, while that proportional to $P^{L}$
describes the longitudinal or color electric sector.  Landau damping provides
an effective infrared cutoff in the magnetic sector, $\sim(\pi
m_{D}^{2}|p_{0}|/4)^{1/3}$, for $|p_{0}|\ll|{\bf p}|$.  This cutoff dominates
the determination of the energy gap; the infrared cutoff given by a putative
magnetic mass \cite{PR3} makes negligible difference.  In the presence of
color pairing, certain of the magnetic gluons acquire a mass as a consequence
of the Meissner effect \cite{HMSW,dirk}, while the Debye mass of some of
the electric gluons is reduced \cite{dirk}.  We ignore here
these effects of the superconducting medium, although they possibly change the
$g$ dependence of the gap matrix in the weak coupling limit.

    For the purpose of solving the gap equation (\ref{phigap}) to leading
logarithmic order in $g$, it is practical to approximate the polarized gluon
propagator (\ref{glueprop}); we first set $-p^2 \approx |{\bf p}|^2$ and
$\chi_L(x)\approx 1$, neglect the real part of $\chi_T(x)$, and replace ${\rm
Im} \chi_T(x) = -\pi x/4$ for small $|x|$ by $-(\pi x/4)
\theta(\sqrt{\pi}m_{D}/2-|{\bf p}|)$.  The effect of this latter replacement
is to introduce a high momentum cutoff for the Landau-damped magnetic gluons.
Then, we replace the gluon propagator by its real part, so that
\begin{equation}
  D_{\mu\nu}^{\alpha\beta}(p)\simeq-{\rm Re}
  \left[
  \frac{\delta_{\alpha\beta}P^{T}_{\mu\nu}}
  {|{\bf p}|^{2}-i\pi m_{D}^{2}p_{0}\theta(\sqrt{\pi}m_{D}/2-|{\bf p}|)
   /4|{\bf p}|}\right]
  -\frac{\delta_{\alpha\beta}P^{L}_{\mu\nu}}
   {|{\bf p}|^{2}+m_{D}^{2}}\ .
  \label{gluePR}
\end{equation}
This form is equivalent to that in Ref.~\cite{PR3}.  In the gap equation
(\ref{phigap}), we may replace $|{\bf q}|$ by $|{\bf k}|$, and then replace
${\bf {\hat q}}$ by ${\bf {\hat k}}$ in the explicit Dirac structures ${\rm
Tr}[\gamma^{\mu}\Lambda^{-} ({\bf {\hat q}})\gamma^{\nu}\Lambda^{+}({\bf {\hat
k}})]P^{L,T}_{\mu\nu} (k-q)$, so that both these terms become $-2$.  We then
find that the gap is determined by
\begin{eqnarray}
  \phi_{+}(\omega_{\pm}({\bf k}),{\bf k}) &=& -\frac{g^{2}}{4}
  \sum_{\eta=\pm}\eta\int\frac{d^{3}q}{(2\pi)^{3}}
  \nonumber \\ & &
  \times\left\{
  \frac{|{\bf q}-{\bf k}|^{2}}
  {|{\bf q}-{\bf k}|^{4}+(\pi m_{D}^{2}(\eta
  \varepsilon_{\eta}({\bf q})
  -\varepsilon_{\pm}({\bf k}))/4|{\bf q}-{\bf k}|)^{2}
   \theta(\sqrt{\pi}m_{D}/2-|{\bf q}-{\bf k}|)}
   \right.
   \nonumber \\ & &
   \left.
    + \frac{1}{|{\bf q}-{\bf k}|^{2}+m_{D}^{2}}
    \right\}
    \nonumber \\ & &
    \times\frac{1}{4}(\lambda^{\alpha})^{T}
    \phi_{+}(\omega_{\eta}({\bf q}),{\bf q})
    \varepsilon^{-1}_{\eta}({\bf q})
    \tanh\left(\frac{\omega_{\eta}({\bf q})}{2T}\right)
    \lambda^{\alpha}\ ,
   \label{phiint}
\end{eqnarray}
where $\theta$ is the angle between ${\bf k}$ and ${\bf q}$.  The dominant
contribution to the integral in Eq.\ (\ref{phiint}) comes from colinear
scattering between the quarks of momenta ${\bf k}$ and ${\bf q}$.
Concentrating on the corresponding momentum region, $\cos\theta\simeq1$ and
$|{\bf k}|\simeq|{\bf q}|$, allows us to solve for the gap to subleading
order in $g$ by reducing Eq.\ (\ref{phiint}) to
\begin{eqnarray}
  \phi_{+}(\omega_{\pm}({\bf k}),{\bf k}) &=& \frac{g^{2}}{16\pi^{2}}
  \sum_{\eta=\pm}\eta\int_{0}^{\infty}d|{\bf q}|
  \left\{
  \frac{1}{3}\ln\left[\frac{|\eta\varepsilon_{\eta}({\bf q})
  -\varepsilon_{\pm}({\bf k})|}{M_{T}}\right]
  +\ln\left(\frac{M_{T}}{2|{\bf q}|}\right)
  +\ln\left(\frac{m_{D}}{2|{\bf q}|}\right)
  \right\}
  \nonumber \\ & &
  \times\frac{1}{4}(\lambda^{\alpha})^{T}
  \phi_{+}(\omega_{\eta}({\bf q}),{\bf q})
  \varepsilon_{\eta}^{-1}({\bf q})
  \tanh\left(\frac{\omega_{\eta}({\bf q})}{2T}\right)
  \lambda^{\alpha}\ ,
   \label{phisimp}
\end{eqnarray}
with $M_{T}\equiv\sqrt{\pi}m_{D}/2$.  Below, we shall consider the two optimal
pairing states analyzed in Sec.\
\ref{sec:GGL} and estimate from Eqs.\ (\ref{na3}) and (\ref{phisimp}) the
chemical potential differences between colors and the reduction in the pairing
gaps induced by imposition of color neutrality.

\subsection{Isoscalar, color-antitriplet channel}
\label{subsec:GE2SC}

    Cooper pairing in the isoscalar phase is described by a gap of the form
(\ref{icat1}), to within a constant phase factor and a special unitary
transformation in flavor space.  This gap matrix is generally accompanied by
nonzero chemical potential differences between colors, as we found in the
Ginzburg-Landau regime in Sec.\ \ref{sec:GGL}.  We estimate here the chemical
potential differences, using the gap equation (\ref{phisimp}) and the color
neutrality conditions (\ref{colsindia}) and (\ref{colsinnondia}) with
(\ref{na3}).

    Note that the right sides of Eq.\ (\ref{na3}) as well as Eq.\
(\ref{phisimp}) with ${\rm Im}{\cal M}=0$ are integrals of an even function of
$\varepsilon({\bf q})$, since under $\varepsilon \to -\varepsilon$, $u_{+}^2
\leftrightarrow v_{+}^2$ and $f \leftrightarrow 1-f$.  It is thus
convenient to diagonalize the matrix $\varepsilon^{2}({\bf q})$ up to first
order in the chemical potential differences, entering via ${\tilde {\cal M}}$.
We first transform bases in color space from the original ($RGB$) basis to a
new basis ($R'G'B'$) to diagonalize the matrix $\phi_{+}^{\dagger}({\bf
q})\phi_{+}({\bf q})$ where $\phi_{+}({\bf q})\equiv\phi_{+}(\varepsilon({\bf
q}),{\bf q})$.  When ${\tilde {\cal M}}=0$, the unitary matrix $U$ that
carries out this diagonalization also diagonalizes $\varepsilon^{2}({\bf q})$.
As we shall see, $U$ is determined solely by the direction of
the vector ${\bf d}$, with components $d_{a}\equiv({\bf d}_{a})_{s}$,
in complex color space and hence is independent of ${\bf q}$.  In
the isoscalar channel where the $ij$-components of $\phi_{+}^{\dagger}
\phi_{+}$ obey $(\phi_{+}^{\dagger}\phi_{+})_{(uu)}=(\phi_{+}^{\dagger}
\phi_{+})_{(dd)}$, $(\phi_{+}^{\dagger}\phi_{+})_{(ss)}=0$, and
$(\phi_{+}^{\dagger}\phi_{+})_{(ij)}=0$ for $i\neq j$, it is sufficient to
diagonalize the submatrix $(\phi_{+}^{\dagger}\phi_{+})_{(uu)}$.  The result
reads
\begin{equation}
 u^{\dagger}(\phi_{+}^{\dagger}\phi_{+})_{(uu)}u=
  \pmatrix{
 |{\bf d}|^{2}  & 0 & 0 \cr
 0 & |{\bf d}|^{2} & 0 \cr
 0 & 0 & 0 \cr
  }\ ,
 \label{submatrix}
\end{equation}
where
\begin{equation}
   u\equiv U_{uu}\equiv({\bf u}_{R'},{\bf u}_{G'},{\bf u}_{B'})\ ,
\end{equation}
with
\begin{equation}
   {\bf u}_{R'}=\frac{{\bf {\hat d}}^{*}{\bf\times}({\bf e\times {\hat d}})}
   {\sqrt{|{\bf e\times {\hat d}}|^{2}}}\ , ~~
   {\bf u}_{G'}=\frac{{\bf e}^{*}{\bf\times}{\bf {\hat d}}^{*}}
   {\sqrt{|{\bf e\times{\hat d}}|^{2}}}\ , ~~
   {\bf u}_{B'}={\bf {\hat d}}\ .
   \label{units}
\end{equation}
Here, ${\bf {\hat d}}\equiv{\bf d}/|{\bf d}|$, and ${\bf e}$ is an
arbitrary complex unit vector that satisfies ${\bf e\times {\hat d}}\neq0$.
Equation (\ref{submatrix}) describes the pairing state with the convention
$d_{R'}=d_{G'}=0$ and $d_{B'}\neq 0$.  The fact that the the matrix
$(\phi_{+}^{\dagger}\phi_{+})_{(uu)}$ is effectively of rank two, 
corresponding to a reduction of the color symmetry from SU(3) to SU(2),
implies that the condensate in the isoscalar, color-antitriplet channel is 
anisotropic in color space.

    The unitary matrix $U$ can be taken to be block diagonal:
$U_{(uu)}=U_{(dd)}=U_{(ss)}=u$ and $U_{(ij)}=0$ for $i\neq j$.  Then,
multiplication of the gap equation (\ref{phisimp}) with ${\rm Im}{\cal M}=0$
by $U$ on the right and $U^{\dagger}$ on the left gives
\begin{eqnarray}
\lefteqn{
u^{\dagger}[\phi_{+}({\bf k})]_{(ud)}u
 \pmatrix{
  0 & 1 & 0 \cr
 -1 & 0 & 0 \cr
  0 & 0 & 0 \cr
}
}
\nonumber \\
&=&\frac{g^{2}}{16\pi^{2}}\int_{0}^{\infty}d|{\bf q}|
\left\{
\frac{1}{3}
\ln\left[{\displaystyle \frac{|u^{\dagger}
\varepsilon^{2}_{(uu)}({\bf k})u
-u^{\dagger}\varepsilon_{(uu)}^{2}({\bf q})u|}{M_{T}^{2}}}
\right]
\pmatrix{
  1 & 0 & 0 \cr
  0 & 1 & 0 \cr
  0 & 0 & 0 \cr
 }
\right.
\nonumber \\  & &
\left.
+\frac{1}{3}
\ln\left[{\displaystyle\frac{|u^{\dagger}
\varepsilon^{2}_{(ss)}({\bf k})u
-u^{\dagger}\varepsilon^{2}_{(ss)}({\bf q})u|}{M_{T}^{2}}}\right]
\pmatrix{
  0 & 0 & 0 \cr
  0 & 0 & 0 \cr
  0 & 0 & 1 \cr
 }
+\ln\left(\frac{M_{T}^{2}}{4|{\bf q}|^{2}}\right)
+\ln\left(\frac{m_{D}^{2}}{4|{\bf q}|^{2}}\right)
\right\}
\nonumber \\  & &
\times\frac{1}{4}u^{\dagger}(\lambda^{\alpha})^{T}u
u^{\dagger}[\phi_{+}({\bf q})]_{(ud)}u
u^{\dagger}\varepsilon^{-1}_{(uu)}({\bf q})
\tanh\left({\displaystyle \frac{\varepsilon_{(uu)}({\bf q})}{2T}}
\right)u
u^{\dagger}\lambda^{\alpha}u
\pmatrix{
  0 & 1 & 0 \cr
 -1 & 0 & 0 \cr
  0 & 0 & 0 \cr
  }\ ,
\label{matrix}
\end{eqnarray}
where $\varepsilon_{(ii)}$ is the $ii$-component of $\varepsilon$, and we
have used $\varepsilon_{(uu)}=\varepsilon_{(dd)}$ and
$\phi_{+(du)}=-\phi_{+(ud)}$.  The Gell-Mann matrix terms can be transformed,
via the Fierz identity $(\lambda^{\alpha})_{ab}(\lambda^{\alpha})_{cd}
=-(2/3)\delta_{ab}\delta_{cd}+2\delta_{ad}\delta_{bc}$, to
\begin{equation}
  \sum_{abcd}u_{ba'}^{*}(\lambda^{\alpha})_{ab}u_{ab'}
  u_{cc'}^{*}(\lambda^{\alpha})_{cd}u_{dd'}=
  -\frac{2}{3}\delta_{a'b'}\delta_{c'd'}
  +2({\bf u}_{b'}{\bf \cdot u}_{d'})
  ({\bf u}_{a'}{\bf \cdot u}_{c'})^{*}\ .
  \label{gm}
\end{equation}

    Before taking into account the effects of nonzero ${\tilde{\cal M}}$, it
is instructive to understand the structure of the gap equation with
${\tilde{\cal M}}=0$.  We first choose the complex unit vector ${\bf e}$ in
such a way as to simplify the matrix $u^{\dagger}\phi_{+(ud)}u$, and then
write down a reduced gap equation.  Because of the
invariance of the grand canonical Hamiltonian under color rotations in the
absence of ${\tilde{\cal M}}$, it is sufficient to treat the states in which
${\bf d\times d}^{*}=0$.  These states are related to the states in which
${\bf d\times d}^{*}\neq0$ by a gauge transformation, ${\bf d}\rightarrow
e^{-2i\varphi}{\bf d}$, and a special unitary transformation, ${\bf d}
\rightarrow U_c {\bf d}$.

   For the states satisfying ${\bf d\times d}^{*}=0$,
we can write ${\bf d}=|{\bf d}|{\bf {\tilde d}}\exp(i\varphi_0)$,
where ${\bf {\tilde d}}$ is the real unit vector proportional to ${\bf d}$,
and $\varphi_0$ is the phase common to all the elements of ${\bf d}$.  Let us
take ${\bf e}$ to be a real unit vector that satisfies ${\bf e\times d}\neq0$.
We then obtain
\begin{equation}
 u^{\dagger}\phi_{+(ud)}u=
 \pmatrix{
 0 & -|{\bf d}| & 0 \cr
 |{\bf d}|\exp(2i\varphi_0) & 0 & 0 \cr
 0 & 0 & 0 \cr
 }\ ,
\end{equation}
as well as the relation, ${\bf u}_{a'}{\bf \cdot u}_{b'}
=\delta_{a'b'}[\delta_{a'R'} +\delta_{a'G'}\exp(-2i\varphi_0)
+\delta_{a'B'}\exp(2i\varphi_0)]$.  The matrix equation (\ref{matrix}) thus
reduces to the single equation,
\begin{eqnarray}
 |{\bf d}({\bf k})|
 &=&-\frac{g^{2}}{24\pi^{2}}
 \int_{0}^{\infty}d|{\bf q}|
 \left\{
 \frac{1}{3}\ln\left(\frac{|E^{2}({\bf q})
 -E^{2}({\bf k})|}{M_{T}^{2}}\right)
 +\ln\left(\frac{M_{T}^{2}}{4|{\bf q}|^{2}}\right)
 +\ln\left(\frac{m_{D}^{2}}{4|{\bf q}|^{2}}\right)
 \right\}
 \nonumber \\ & &
 \times
 |{\bf d}({\bf q})|
 E^{-1}({\bf q})\tanh\left(\frac{E({\bf q})}{2T}\right)\ ,
  \label{deq}
\end{eqnarray}
where
\begin{equation}
   E({\bf q})= \left[\left(|{\bf q}|-\frac{\mu}{3}\right)^{2}+
   |{\bf d}({\bf q})|^{2}\right]^{1/2}
  \label{Eofq}
\end{equation}
is the $R'R'$ and $G'G'$ component of the diagonal matrix
$\left([u^{\dagger}\varepsilon^{2}_{(uu)} ({\bf q})u]_{\tilde{\cal
M}=0}\right)^{1/2}$.  Note that Eq.\ (\ref{deq}) with $d_{R}=d_{G}=0$ is
equivalent to the gap equation usually analyzed for the isoscalar
color-antitriplet channel (see, e.g., Refs.\ \cite{PR3,SW}).  In the weak
coupling limit of interest here, where a tiny energy gap arises in the
momentum region immediately close to the Fermi surface, we can replace the
$E^{2}({\bf x})$ $({\bf x} ={\bf k},{\bf q})$ in the logarithmic term of Eq.\
(\ref{deq}) with the $B'B'$-component $(|{\bf x}|-\mu/3)^{2}$ of the diagonal
matrix.  We then assume in Eq.\ (\ref{deq}) that $|{\bf q}|\approx\mu/3$ in 
the last two logarithms and shift the integration variable from $|{\bf q}|$ to
$|{\bf q}|-\mu/3$.  The resulting equation reads
\begin{equation}
  |{\bf d}({\bf k})|
  =\frac{g^{2}}{18\pi^{2}}
  \int_{0}^{\delta}\frac{d(|{\bf q}|-\mu/3)}{E({\bf q})}
  \tanh\left(\frac{E({\bf q})}{2T}\right)
  \frac{1}{2}
  \ln\left[\frac{(b\mu/3)^{2}}{|(|{\bf q}|-\mu/3)^{2}
  -(|{\bf k}|-\mu/3)^{2}|}\right]|{\bf d}({\bf q})|\ ,
  \label{deqf}
\end{equation}
with
\begin{equation}
  b\equiv256\pi^{4}\left(\frac{2}{3g^{2}}\right)^{5/2} ;
\end{equation}
the cutoff $\delta$, obeying $|{\bf d}(|{\bf q}|=\mu/3)|\ll\delta\ll
m_{D}$, is chosen so that $|{\bf d}(|{\bf q}|>\delta)|$ is vanishingly small.
Here, we have used $m_{D}^{2}=g^2 \mu^{2}/6\pi^{2}+{\cal O}(g^{2}T^{2})$.

    Let us now consider a color-singlet system and ask, in particular, whether
or not the degeneracy of the order-parameter sets is removed in the weak 
coupling limit by the chemical potential differences between colors stemming 
from color neutrality (\ref{colsindia}) and (\ref{colsinnondia}).  As a first
step, we write the solution to the gap equation (\ref{deqf}), following a line
of argument of Pisarski and Rischke \cite{PR3}, and express the chemical 
potential deviations ${\tilde\mu}_{ab}$ in terms of the obtained gap.  We then
examine how such deviations in turn affect the gap itself.  As we shall see, 
the degeneracy is not removed up to subleading order in $g$.

    We now implement a process developed by Pisarski and Rischke \cite{PR3} to
convert the integral equation (\ref{deqf}) into an equivalent differential
equation [Eq.\ (\ref{diffeq}) below] via the approximation,
\begin{eqnarray}
  \frac{1}{2}\ln\left[\frac{(b\mu/3)^{2}}{|(|{\bf q}|-\mu/3)^{2}
 -(|{\bf k}|-\mu/3)^{2}|}\right]
 &\rightarrow&
  \theta(||{\bf q}|-\mu/3|-||{\bf k}|-\mu/3|)
\ln\left(\frac{b\mu/3}{||{\bf q}|-\mu/3|}\right)
 \nonumber \\ & &
 +\theta(||{\bf k}|-\mu/3|-||{\bf q}|-\mu/3|)
\ln\left(\frac{b\mu/3}{||{\bf k}|-\mu/3|}
 \right)\ .
\end{eqnarray}
With this approximation and the change of variables,
\begin{equation}
  x\equiv\ln\left[\frac{2b\mu/3}{||{\bf k}|-\mu/3|+E({\bf k})}\right]\ , ~~
  y\equiv\ln\left[\frac{2b\mu/3}{||{\bf q}|-\mu/3|+E({\bf q})}\right]\ , ~~
  x_{F}\equiv\ln\left[\frac{2b\mu/3}{E(|{\bf k}|=\mu/3)}\right]\ ,
  \label{xy}
\end{equation}
we obtain
\begin{equation}
  |{\bf d}(x)|={\bar g}^{2}\left\{x\int_{x}^{x_{F}}dy
    \tanh\left(\frac{E(y)}{2T}\right)|{\bf d}(y)|
    +\int_{\ln(b\mu/3\delta)}^{x}dy y
    \tanh\left(\frac{E(y)}{2T}\right)|{\bf d}(y)|
\right\}\ ,
  \label{intstep}
\end{equation}
where ${\bar g}\equiv g/3\sqrt{2}\pi$, and
$\ln(b\mu/3||{\bf k}|-\mu/3|)\simeq x$ and $\ln(b\mu/3||{\bf q}|-\mu/3|)
\simeq y$ have been used.  Differentiation of
Eq.\ (\ref{intstep}) with respect to $x$ leads finally to
\begin{equation}
  \frac{d^{2}|{\bf d}(x)|}{dx^{2}}=-{\bar g}^{2}
  \tanh\left(\frac{E(x)}{2T}\right)|{\bf d}(x)|\ .
\label{diffeq}
\end{equation}

    We summarize the results of Eq.\ (\ref{diffeq}), as obtained by Pisarski
and Rischke \cite{PR3}.  At $T=0$, the magnitude of the gap is given by
\begin{equation}
    |{\bf d}(x)|=\frac23 b\mu e^{-\pi/2{\bar g}} \sin({\bar g} x)\ .
  \label{prgap}
\end{equation}
It was pointed out that the BCS-like exponential term and the sinusoidal
$x$ dependence arise from nearly static magnetic gluons that undergo Landau
damping and mediate the long-range part of the magnetic interactions, and that
both the higher frequency magnetic gluons, which are little affected by Landau
damping, and Debye-screened electric gluons play a dominant role in
determining the pre-exponential factor.

    The overall coefficient of the $\mu/g^{5}$ in the prefactor is usually
considered to be correct up to a factor of order unity, for the reason that
radiative corrections such as quasiparticle wave function renormalization
\cite{son} and vertex corrections \cite{SW} do not modify the asymptotic form
of the gap.  However, it is still uncertain the extent to which this result is
exact since contributions of the Meissner effect and Debye screening in the
superconducting medium remain to be clarified \cite{dirk}.

    Expression (\ref{prgap}) indicates that the gap takes on a peak value
$|{\bf d}(x_{F})|=(2b\mu/3)e^{-\pi/2{\bar g}}$ at $|{\bf k}|=\mu/3$,
smoothly reaches half its peak value at $||{\bf k}|-\mu/3|\sim |{\bf
d}(x_{F})|e^{\pi/3{\bar g}}$, and becomes of order $(g\ln g)|{\bf
d}(x_{F})|$ at $||{\bf k}|-\mu/3|\sim m_{D}$.
The critical temperature $T_{c}$ is given by
\begin{equation}
   T_{c}=\frac{e^{\gamma}}{\pi}|{\bf d}(x_{F},T=0)|\ ,
  \label{Tc}
\end{equation}
where $\gamma=0.5772\ldots$ is the Euler constant.  It was stressed by
Pisarski and Rischke \cite{PR3} that expression (\ref{Tc}) is of the usual BCS
form, a feature stemming from the fact that in the weak coupling limit
\cite{note1}, $|{\bf d}(x,T)|\simeq|{\bf d}(x,T=0)|(|{\bf d}(x_{F},T)|
/|{\bf d}(x_{F},T=0)|)$.  This expression, except for a factor of order unity
due to quasiparticle wave function renormalization, was reproduced by Brown,
Liu, and Ren \cite{BLR} using finite temperature diagrammatic perturbation
theory in the normal phase.  Since ${\tilde{\cal M}}$ vanishes at $T=T_{c}$,
the critical temperature itself is unchanged by the color neutrality
constraint.

    We turn to the calculations of the chemical potential deviations ${\tilde
\mu}_{ab}$ as functions of $|{\bf d}|$ given by Eq.\ (\ref{prgap}); we consider
only zero temperature for simplicity.  Up to lowest order in ${\tilde{\cal
M}/\mu}$, the difference between $n_{ab}$ given by Eq.\ (\ref{na3}) and the
corresponding normal-phase value $\delta_{ab}n_{\rm N}\equiv\delta_{ab}
2(N_{f}/3)N(\mu/3)\mu/3$ can be written as
\begin{equation}
  n_{ab}-n_{\rm N}\delta_{ab}=2N_{f}N(\mu/3){\tilde{\mu}}_{ab}
  +2\sum_{a'=R',G'}u_{aa'}u_{ba'}^{*}\delta n\ ,
  \label{nadif}
\end{equation}
where
\begin{equation}
  \delta n\equiv \frac1V
     \left\{\sum_{|{\bf q}|<\mu/3}
  \left[-1-\frac{|{\bf q}|-\mu/3}{E({\bf q})}\right]
  +\sum_{|{\bf q}|>\mu/3}
  \left[1-\frac{|{\bf q}|-\mu/3}{E({\bf q})}\right]
   \right\}
  \label{deltan}
\end{equation}
is the number density excess due to pairing \cite{note2}, and $V$ is the
system volume.  This excess can be estimated from Eqs.\ (\ref{Eofq}) and
(\ref{prgap}) as
\begin{equation}
 \delta n\simeq \frac{3\pi}{2{\bar g}\mu}
  N(\mu/3)|{\bf d}(x_{F})|^{2}\ .
\end{equation}
Using these expressions in the color neutrality conditions
(\ref{colsindia}) and (\ref{colsinnondia}), we obtain, with the help of
Eqs.\ (\ref{deltamu2}) and (\ref{units}),
\begin{equation}
  {\tilde\mu}_{ab}=(3{\hat d}_{a}{\hat d}_{b}^{*}-\delta_{ab})
   \frac{\delta n}{9N(\mu/3)}\ .
  \label{muazero}
\end{equation}
The chemical potential deviations ${\tilde\mu}_{ab}$ are thus of order
$|{\bf d}(x_{F})|^{2}/{\bar g}\mu$.

    Up to subleading order in $g$, such chemical potential deviations do not
modify any of the factors in Eq.~(\ref{prgap}) for the magnitude of the gap
$|{\bf d}|$.  This is because $||{\bf q}|-\mu/3||{\tilde\mu}_{ab}|\ll |{\bf
d}(x_{F})|^{2}$ in the momentum region, $||{\bf q}|-\mu/3|\lesssim \delta$,
where the gap is appreciable.  We remark that this robustness of the
exponential term and the leading pre-exponential factor in Eq.\ (\ref{prgap})
is supported by the renormalization-group analysis developed by Son
\cite{son}.  For example, let us consider the case in which $d_{R}=d_{G}=0$
and $d_{B}\neq0$.  In this case, Cooper pairing occurs between the quarks of
color $R$ and $G$ with momenta $\pm{\bf k}$ close to the Fermi surface having
the Fermi momentum $k_{F}=\mu/3+{\tilde\mu}_{RR}=\mu/3+{\tilde\mu_{GG}}$.  In
the renormalization-group analysis, the scattering amplitude between these
quarks is characterized by the parameter $t=-\ln||{\bf k}|-k_{F}|$.  The
scattering amplitude becomes singular, as Gorkov originally showed, at the
onset of pairing.  The value of $t$ at which this singularity occurs -- the
Landau pole $t_{L}$ -- tells us the scale of the energy gap according to
$|{\bf d}(|{\bf k}|=k_{F})|\sim m_{D}e^{-t_{L}}$.  Up to subleading order in 
$g$, non-zero ${\tilde\mu}_{RR}$ has no effect on $t_{L}$, which behaves as 
$t_{L}=\pi/2{\bar g}+6\ln g+\cdots$.

    In the weak coupling limit, up to subleading order in $g$, we thus find
that the degenerate set of ${\bf d}$'s obtained with ${\tilde{\cal M}}=0$
persists in the color-singlet system studied here, a result consistent with
that obtained from the general Ginzburg-Landau theory in Sec.\
\ref{sec:GGL}.  Nevertheless, it is instructive to examine the leading
contribution of  ${\tilde\mu}_{ab}$ to the zero-temperature gap, in view of
the fact that a strongly coupled quark system may develop a considerable
energy gap (see, e.g., Ref.\ \cite{instanton}).

    The properties of this contribution can be roughly understood by finding
the eigenvalues of $\varepsilon^{2}_{(uu)}({\bf q})$,
$=\varepsilon^{2}_{(dd)}({\bf q})$, the square of the quasiparticle energy
given by Eq.\ (\ref{qpenergy}).  This analysis is only applicable to the case
in which ${\bf d}\times{\bf d}^{*}=0$ and hence ${\rm Im}M=0$.  (If ${\rm
Im}M\neq0$, one must go back to the original gap equation (\ref{phisimp})
dependent on the matrices $\varepsilon_{\eta}$ and $\omega_{\eta}$, which are
not diagonalized by the same unitary transformation.)  Up to ${\cal
O}({\tilde{\cal M}})$, these eigenvalues are given by
\begin{eqnarray}
  & &E_{R'}^{2}({\bf q})=
  \left(|{\bf q}|-\frac{\mu}{3}-\delta\mu_{R'}\right)^{2}+
  |{\bf d}({\bf q})|^{2}\ ,~~
  \nonumber \\ & &
  E_{G'}^{2}({\bf q})=
  \left(|{\bf q}|-\frac{\mu}{3}-\delta\mu_{G'}\right)^{2}+
  |{\bf d}({\bf q})|^{2}\ ,~~
  \nonumber \\ & &
  E_{B'}^{2}({\bf q})=\left(|{\bf q}|-\frac{\mu}{3}-
  \delta\mu_{B'}\right)^{2}\ ,
  \label{eigen}
\end{eqnarray}
where
\begin{equation}
  \delta\mu = \delta\mu_{B'} = -2\delta\mu_{R'}=
  -2\delta\mu_{G'}
  \equiv\frac{\pi|{\bf d}(x_{F})|^{2}}{3{\bar g}\mu}
  \label{defdeltamu}
\end{equation}
is the chemical potential shift from $\mu/3$.  The order of the
eigenvalues corresponds to that in Eq.\ (\ref{submatrix}).  Note that the
chemical potential shift $\delta\mu$, stemming from the anisotropy of the
condensate in color space, is independent of its color orientation.
This reflects the fact that the properties of a
color-singlet system do not depend on the choice of color axes.  It is also
important to note that $\delta\mu$ is positive definite.  This feature arises
due to the property discussed in note \cite{note2}.  Not only does
positive $\delta\mu$ enlarge the Fermi surface of the gapless quarks of color
$B'$, but it also decreases the Fermi energy of the quarks of colors $R'$ or
$G'$, leading to overall suppression of the energy gap.

    In order to calculate in detail the ${\cal O}({\tilde\mu}_{ab})$
correction to the magnitude of the gap in the case in which
${\bf d}\times{\bf d}^{*}=0$, we note
that $U$ diagonalizes ${\cal M}$ and hence $\varepsilon^{2}({\bf q})$.  This
is evident from the relation,
\begin{equation}
 {\cal M}_{abij}=\delta_{ij}\left\{\left[\frac{\mu}{3}+\frac
 {\pi|{\bf d}(x_{F})|^{2}}{3{\bar g}\mu}\right]\delta_{ab}
 -\frac{\pi[(\phi_{+}^{\dagger}
 \phi_{+})_{(uu)}]_{|{\bf k}|=\mu/3}}{2{\bar g}\mu}\right\}\ ,
\end{equation}
obtained from Eq.\ (\ref{muazero}).  It is thus straightforward to show that
the gap equation (\ref{matrix}) reduces to Eq.\ (\ref{deq}) in which the
quasiparticle energy $E({\bf q})$ is replaced by $E_{R'}({\bf q})$,
=$E_{G'}({\bf q})$.  The solution to this equation reads
\begin{eqnarray}
 |{\bf d}(x')|&=&\frac{2C_{1}b\mu}{3}
 \exp\left(-\frac{\pi}{2{\bar g}}\right)\sin({\bar g} x')\ ,
 \nonumber \\
 x'&\equiv&\ln\left[\frac{2C_{1}b\mu/3}{||{\bf k}|-\mu/3
 +\delta\mu/2|+E_{R'}({\bf k})}\right]\ ,
\end{eqnarray}
where the correction factor is
\begin{equation}
  C_{1}\simeq1-9\frac{\delta\mu}{\mu}\ ,
\end{equation}
and we have used $m_{D}^{2}=g^2\mu^2/6\pi^2+{\cal O}(g^2{\delta\mu}^2)$.

    Note that $C_{1}$ is less than unity.  As a consequence the anisotropy of
the condensate in a color-singlet system acts to reduce the gap size.

    This reduction, together with the reduction in the densities of states
at the Fermi surfaces for colors $R'$ and $G'$ due to $\delta\mu_{R',G'}$,
leads to a decrease in the condensation energy of the system.  These features
arise despite the fact that there are no Fermi momentum
differences between the two colors associated with the pairing.  This
situation is in contrast to the case of ferromagnetic superconductors
\cite{FFLO} and to the case of color-flavor-locked quark superfluids with
nonzero strange quark mass \cite{mass}, where Fermi momentum mismatches
between the electron spins and between the quark flavors involved in the
pairing, respectively, play a role in reducing the magnitude of the gap.

\subsection{Color-flavor locking}
\label{subsec:GECFL}

    We proceed to examine color-flavor locking, described by the order
parameter (\ref{phicfl}) to within a constant phase factor and
a special unitary transformation in flavor space.
We first note the relation ${\tilde\mu}_{ab}
\propto\sum_{i}[(\phi_{+}^{\dagger}\phi_{+})_{|{\bf k}|=\mu/3}]_{abii}+{\rm
const.}\times\delta_{ab}$, which can be derived up to first order in
$\phi_{+}^{\dagger}\phi_{+}$ from the color-singlet conditions
(\ref{colsindia}) and (\ref{colsinnondia}) as well as Eq.\ (\ref{na3}).  Here,
the constant affixed to $\delta_{ab}$ ensures Eq.\ (\ref{deltamu2}).
Substituting Eq.\ (\ref{phicfl}) into this relation we find
$\sum_{i}[(\phi_{+}^{\dagger}\phi_{+})_{|{\bf k}|=\mu/3}]_{abii}
=4\delta_{ab}|\kappa_{A}|^{2}$, and thus obtain
${\tilde{\cal M}}=0$.  This result, coming from the isotropy of the condensate
in color space, is consistent with that from the general Ginzburg-Landau
analysis in Sec.\ \ref{sec:GGL}.  We can
thus derive the behavior of the energy gap at $T=0$ and near $T_c$ from the
gap equation (\ref{phisimp}) with ${\tilde{\cal M}}=0$.  Substitution of
ansatz (\ref{phicfl}) and ${\tilde{\cal M}}=0$ into Eq.~(\ref{phisimp})
leads to an equation for $\kappa_{A}$,
\begin{eqnarray}
 \kappa_{A}({\bf k})
 &=&\frac{g^{2}}{18\pi^{2}}
 \int_{0}^{\delta}d(|{\bf q}|-\mu/3)
 \frac{1}{2}
 \ln\left[\frac{(b\mu/3)^{2}}{|(|{\bf q}|-\mu/3)^{2}
 -(|{\bf k}|-\mu/3)^{2}|}\right]
 \nonumber \\ & &
 \times\left\{-\frac{2}{3}
 \left[\frac{-\kappa_{A}({\bf q})}
 {E_{(8)}({\bf q})}\right]
 \tanh\left(\frac{E_{(8)}({\bf q})}{2T}\right)
 \right.
 \nonumber \\ & &
 \left.
 +\frac{1}{6}
 \left[\frac{2\kappa_{A}({\bf q})}
 {E_{(1)}({\bf q})}\right]
  \tanh\left(\frac{E_{(1)}({\bf q})}{2T}\right)\right\}\ ,
 \label{KA}
\end{eqnarray}
with
\begin{eqnarray}
 & &E_{(8)}({\bf q})= \left((|{\bf q}|-\mu/3)^{2}+
 |\kappa_{A}({\bf q})|^{2}\right)^{1/2} \ , ~~
 \nonumber \\ & &
 E_{(1)}({\bf q})= \left( (|{\bf q}|-\mu/3)^{2}+
 4|\kappa_{A}({\bf q})|^{2}\right)^{1/2}\ .
\end{eqnarray}
Here, we have used the color structure,
\begin{equation}
  \sum_{bc}
  \frac{1}{4}
  (\lambda^{\alpha})_{ba}(\delta_{bi}\delta_{cj}-\delta_{bj}\delta_{ci})
  (\lambda^{\alpha})_{cd}=-\frac{2}{3}
  (\delta_{ai}\delta_{dj}-\delta_{aj}\delta_{di})\ ,
\end{equation}
and the diagonalized form of the quasiparticle energy squared,
\begin{equation}
  U_{\rm CFL}^{T}\varepsilon^{2}({\bf q})U_{\rm CFL}
  =\delta_{a'b'}\delta_{i'j'}E^{2}_{(8)}({\bf q})
  +\delta_{a'B'}\delta_{b'B'}\delta_{i's'}\delta_{j's'}
  [E^{2}_{(1)}({\bf q})-E^{2}_{(8)}({\bf q})]\ ,
\end{equation}
with
\begin{equation}
U_{\rm CFL}=\frac{1}{\sqrt{6}}
\pmatrix{
-2&0&0&0&0&0&0&0&\sqrt{2} \cr
0&\sqrt{6}&0&0&0&0&0&0&0 \cr
0&0&\sqrt{6}&0&0&0&0&0&0 \cr
0&0&0&\sqrt{6}&0&0&0&0&0 \cr
1&0&0&0&-\sqrt{3}&0&0&0&\sqrt{2} \cr
0&0&0&0&0&\sqrt{6}&0&0&0 \cr
0&0&0&0&0&0&\sqrt{6}&0&0 \cr
0&0&0&0&0&0&0&\sqrt{6}&0 \cr
1&0&0&0&\sqrt{3}&0&0&0&\sqrt{2} \cr
}\ ;
\end{equation}
here the bases are taken to be $(Ru,Gu,Bu,Rd,Gd,Bd,Rs,Gs,Bs)$ in the
original color-flavor space and
$(R'u',G'u',B'u',R'd',G'd',B'd',R's',G's',B's')$ in the transformed
color-flavor space.

    At $T=0$, the gap equation (\ref{KA}) is equivalent to that obtained by
Sch{\"a}fer \cite{tom}.  The analogy between Eq.\ (\ref{KA}) and Eq.\
(\ref{deqf}) allows us to write $\kappa_{A}$ in the form (\ref{prgap}) valid
up to subleading order in $g$ \cite{tom}:
\begin{equation}
 \kappa_{A}(x_{(8)})  =2^{-1/3}\exp(i\varphi_A)
  \frac{2b\mu}{3}\exp\left(-\frac{\pi}{2{\bar g}}\right)
  \sin({\bar g} x_{(8)})\ ,
  \label{KA1}
\end{equation}
where
\begin{equation}
  x_{(8)}\equiv\ln\left(\frac{2^{2/3}b\mu/3}{||{\bf k}|-\mu/3|
  +E_{(8)}({\bf k})}\right)\ ,
\end{equation}
and $\varphi_A$ is the phase of $\kappa_{A}$.

    At finite temperatures, the assumption that $\kappa_{A}({\bf k},T) \simeq
\kappa_{A}({\bf k},T=0)(\kappa_{A}(|{\bf k}|=\mu/3,T) /\kappa_{A}(|{\bf
k}|=\mu/3,T=0))$ holds as in the isoscalar, color-antitriplet channel.  At 
$T=T_c$, where $E_{(8)}({\bf q})=E_{(1)}({\bf q})$, the gap equation 
(\ref{KA}) has the same structure as Eq.\ (\ref{deqf}) obtained for the 
isoscalar, color-antitriplet channel.  Thus, the critical temperature $T_{c}$
equals the result, Eq.\ (\ref{Tc}), in the isoscalar channel.  Such equality,
persisting in the strong coupling regime (see Sec.\ \ref{sec:GGL}), stems from
the fact that both these types of pairing are induced by the same instability 
of the normal phase.

\section{Ginzburg-Landau region at high densities}
\label{sec:GL}

    We are now in a position to calculate, in the weak coupling limit, the
Ginzburg-Landau free energy, (\ref{deltaomega}), the difference between the
superfluid and normal phases near the transition temperature $T_{c}$.  As
earlier, we ignore the normal state Hartree-Fock terms in the Schwinger-Dyson
equation (\ref{self}), and identify the thermodynamic potential in the normal
phase with that of an ultrarelativistic, noninteracting Fermi gas of Fermi
energy $\mu/3$.  We thus obtain \cite{FM}
\begin{equation}
 \Delta\Omega=\Delta\Omega_{\rm ideal}+\Omega_{\rm loop}+
 \Omega_{\rm res}\ ,
 \label{deltawc}
\end{equation}
where
\begin{equation}
 \Delta\Omega_{\rm ideal}
 =-\frac{1}{12\pi^{2}}\left[{\rm Tr}{\cal M}^{4}
  -3N_{f}\left(\frac{\mu}{3}\right)^{4}\right]
 \label{ideal}
\end{equation}
is the difference in the ideal-gas contribution between the superfluid and
normal phases;
\begin{equation}
 \Omega_{\rm loop}=-\frac{T}{2}\sum_{n~ {\rm odd}}
 \int\frac{d^{3}q}{(2\pi)^{3}}
 {\rm Tr}[-G(q)\Sigma(q)+\ln G_{0}^{-1}(q)G(q)]
  \label{loop}
\end{equation}
is the contribution, again up to ${\cal O}(g^{2})$, of quark loops in the
superfluid vacuum; and
\begin{equation}
 \Omega_{\rm res}=-\left\{\frac{T}{4}\sum_{n~ {\rm odd}}
 \int\frac{d^{3}q}{(2\pi)^{3}}
 {\rm Tr}[G(q)\Sigma(q)]\right\}_{T=T_{c},{\tilde{\cal M}}=0}\ ,
  \label{omegares}
\end{equation}
resulting from the mean-field approximation \cite{PR1} adopted in
writing Eq.\ (\ref{self}), guarantees that the energy gap $\Delta$ vanishes at
$T=T_{c}$.

    We now expand the thermodynamic potential difference (\ref{deltawc}) with
respect to $\Delta({\bf q})\equiv\Delta(q_{0}=|{\bf q}|-\mu/3,{\bf q})$ and
${\tilde \mu}_{ab}$.  In calculating $\Omega_{\rm loop}$ up to ${\cal O}
(\Delta^{4})$ and ${\cal O}(\Delta^{2}{\tilde {\cal M}})$, we use the
quark propagator $G$ given by Eq.\ (\ref{G}), the proper self energy $\Sigma$
given by Eq.\ (\ref{GSig}), and the noninteracting quark propagator $G_{0}$
given by Eq.\ (\ref{G0}).

    The explicit Dirac structure of $G$ can be obtained from $G^{(11)}$,
Eq.~(\ref{G110}), $G^{(21)}$, Eq.~(\ref{G21}),
\begin{equation}
  G^{(12)}(q)=
  \left[{\tilde \Delta}(q)(\gamma q-\gamma^{0}{\cal M}^{T})^{-1}\Delta(q)
  -(\gamma q+\gamma^{0}{\cal M})\right]^{-1}
  {\tilde \Delta}(q)
  (\gamma q-\gamma^{0}{\cal M}^{T})^{-1}\ ,
\end{equation}
and
\begin{equation}
  G^{(22)}(q)=
  \left[(\gamma q-\gamma^{0}{\cal M}^{T})
  -\Delta(q)(\gamma q+\gamma^{0}{\cal M})^{-1}{\tilde\Delta}(q)
  \right]^{-1}\ .
\end{equation}
Thus,
\begin{eqnarray}
  {\rm Tr}[G(q)\Sigma(q)]&=&
  2{\rm Tr}\left\{\frac{1}{\gamma q+\gamma^{0}{\cal M}}
  {\tilde\Delta}(q)\frac{1}{\gamma q-\gamma^{0}{\cal M}^{T}}\Delta(q)
  \right.
 \nonumber \\ & &
  \left.
  +\left[\frac{1}{\gamma q+\gamma^{0}{\cal M}}
  {\tilde\Delta}(q)\frac{1}{\gamma q-\gamma^{0}{\cal M}^{T}}\Delta(q)
  \right]^{2}+\cdots
  \right\}\ ,
   \label{loopa}
\end{eqnarray}
and
\begin{eqnarray}
  {\rm Tr}[\ln G_{0}^{-1}(q)G(q)]&=&
  {\rm Tr}\left\{\frac{1}{\gamma q+\gamma^{0}{\cal M}}
  {\tilde\Delta}(q)\frac{1}{\gamma q-\gamma^{0}{\cal M}^{T}}\Delta(q)
  \right.
 \nonumber \\ & &
  \left.
  +\frac{1}{2}\left[\frac{1}{\gamma q+\gamma^{0}{\cal M}}
  {\tilde\Delta}(q)\frac{1}{\gamma q-\gamma^{0}{\cal M}^{T}}\Delta(q)
  \right]^{2}+\cdots
 \right\}\ .
 \label{loopb}
\end{eqnarray}
Combining Eqs.\ (\ref{loopa}) and (\ref{loopb}) with (\ref{loop}) and
performing the Matsubara frequency summation, we finally obtain
\begin{equation}
 \Omega_{\rm loop}= \Omega_{\rm loop}^{(0)}+\Omega_{\rm loop}^{(1)}\ ,
 \label{loop0}
\end{equation}
where
\begin{eqnarray}
  \Omega_{\rm loop}^{(0)}&=&
  -\frac{1}{2}\int\frac{d^{3}q}{(2\pi)^{3}}
  \tanh\left(\frac{|{\bf q}|-\mu/3}{2T}\right)
  \frac{1}{|{\bf q}|-\mu/3}
  {\rm Tr}[\phi^{\dagger}_{+}({\bf q})\phi_{+}({\bf q})]
 \nonumber \\ & &
  +\frac{3}{8}\int\frac{d^{3}q}{(2\pi)^{3}}
  \left[\tanh\left(\frac{|{\bf q}|-\mu/3}{2T}\right)
  \frac{1}{(|{\bf q}|-\mu/3)^3}\right.
\nonumber \\  & &
  -\left.\cosh^{-2}\left(\frac{|{\bf q}|-\mu/3}{2T}\right)
   \frac{1}{2T(|{\bf q}|-\mu/3)^{2}}\right]
\nonumber \\  & &
  \times{\rm Tr}[\phi^{\dagger}_{+}({\bf q})\phi_{+}({\bf q})
  \phi^{\dagger}_{+}({\bf q})\phi_{+}({\bf q})]
 \label{loop00}
\end{eqnarray}
is the usual Ginzburg-Landau expansion obtained for ${\tilde {\cal M}}=0$,
and
\begin{eqnarray}
 \Omega_{\rm loop}^{(1)} &=&
 -\frac{1}{2}\int\frac{d^{3}q}{(2\pi)^{3}}\left[
 \tanh\left(\frac{|{\bf q}|-\mu/3}{2T}\right)\frac{1}{(|{\bf q}|-\mu/3)^{2}}
 \right.
\nonumber \\ & &
 -\left.\cosh^{-2}\left(\frac{|{\bf q}|-\mu/3}{2T}\right)
   \frac{1}{2T(|{\bf q}|-\mu/3)}\right]
 {\rm Tr}[\phi^{\dagger}_{+}({\bf q})\phi_{+}({\bf q}){\tilde {\cal M}}]
 \label{loop01}
\end{eqnarray}
is the ${\cal O}({\tilde {\cal M}})$ correction to the first term on the
right side of Eq.\ (\ref{loop00}).  We have again ignored the contributions of
antiquark quasiparticles, and used the relation ${\rm Tr}(\phi^{\dagger}_{+}
\phi_{+}{\tilde {\cal M}})= {\rm Tr}(\phi_{+}\phi_{+}^{\dagger}{\tilde {\cal
M}}^{T})$, coming from $\phi_{+}^{T}({\bf q})=\phi_{+}({\bf q})$ [see Eq.\
(\ref{pauli1})].

    We turn now to calculating $\Omega_{\rm loop}^{(0)}$ and $\Omega_{\rm
loop}^{(1)}$ for the isoscalar, color-antitriplet channel considered in Secs.\
\ref{sec:GGL} and \ref{sec:GE}, and derive the high-density Ginzburg-Landau
form of $\Delta\Omega$ used in Sec.\ \ref{sec:GGL}.  We remark in passing that
the calculations of $\Omega_{\rm loop}^{(0)}$ for the color-flavor locking
reproduce the values of ${\bar\alpha}$, $\beta_{1}$, and $\beta_{2}$ that will
be derived below for the isoscalar channel.

    To derive the coefficients of the Ginzburg-Landau free energy for
isoscalar, color-antitriplet pairing, we first carry out the integrals over
${\bf q}$ in $\Omega_{\rm loop}^{(0)}$, Eq.\ (\ref{loop00}), and $\Omega_{\rm
loop}^{(1)}$, Eq.\ (\ref{loop01}).  Here we evaluate $\phi_{+}({\bf q},T)$,
whose ${\bf q}$ and $T$ dependences are effectively decoupled for
$T\leq T_{c}$ (see note \cite{note1}), as
\begin{equation}
 \phi_{+}({\bf q},T) \simeq \phi_{+}(|{\bf q}|=\mu/3,T)
    \sin({\bar g}y)|_{T=0}\ ,
 \label{approx1}
\end{equation}
where $y$ is given by Eq.\ (\ref{xy}).  We also use the approximation,
\begin{eqnarray}
 \tanh\left(\frac{|{\bf q}|-\mu/3}{2T}\right)
 &\rightarrow&\theta(\kappa|{\bf d}(x_{F},T=0)|-||{\bf q}|-\mu/3|)
 \tanh\left(\frac{|{\bf q}|-\mu/3}{2T}\right)
 \nonumber \\ & &
 +\theta(||{\bf q}|-\mu/3|-\kappa|{\bf d}(x_{F},T=0)|)
 \frac{|{\bf q}|-\mu/3}{||{\bf q}|-\mu/3|}\ ,
 \label{approx2}
\end{eqnarray}
where $\kappa$ is a positive number of order $g^{-1}$, and $x_{F}$ is
given by Eq.\ (\ref{xy}).  With this choice, $\tanh((|{\bf
q}|-\mu/3)/2T)\simeq (|{\bf q}|-\mu/3)/||{\bf q}|-\mu/3|$ for $||{\bf
q}|-\mu/3|\gtrsim\kappa|{\bf d}(x_{F},T=0)|$, and $\sin({\bar
g}y)|_{T=0}\simeq1$ for $||{\bf q}|-\mu/3|\lesssim\kappa|{\bf
d}(x_{F},T=0)|$.  We thus obtain
\begin{eqnarray}
 \Omega_{\rm loop}^{(0)}&=&-N(\mu/3)
 \ln\left[\frac{e^{\gamma}(2b\mu/3)^{1/2}|{\bf d}(x_{F},T=0)|^{1/2}}
 {\pi T}\right]{\rm Tr}(\phi_{+}^{\dagger}\phi_{+})_{F}
 \nonumber \\ & &
 +\frac{3}{2}\frac{7\zeta(3)}{8(\pi T_{c})^{2}}N(\mu/3)
 {\rm Tr}(\phi_{+}^{\dagger}\phi_{+}\phi_{+}^{\dagger}\phi_{+})_{F}\ ,
 \label{omegaloop0}
\end{eqnarray}
and
\begin{equation}
 \Omega_{\rm loop}^{(1)}=\frac{3}{\mu}\ln\left(\frac{3T_{c}}{\mu}\right)
  N(\mu/3)
  {\rm Tr}(\phi_{+}^{\dagger}\phi_{+}{\tilde {\cal M}})_{F}
  \equiv \chi
  {\rm Tr}(\phi_{+}^{\dagger}\phi_{+}{\tilde {\cal M}})_{F} \ ,
 \label{chi1}
\end{equation}
where the coefficients of ${\rm Tr}(\cdots)_{F}$ include the leading
contributions with respect to $g$, and the temperature has been set equal to
the critical temperature Eq.\ (\ref{Tc}), except in the coefficient of ${\rm
Tr}(\phi_{+}^{\dagger}\phi_{+})_{F}$.  The term $\Omega_{\rm res}$,
expressed as
\begin{eqnarray}
 \Omega_{\rm res} &=& N(\mu/3)
 \ln\left[\frac{e^{\gamma}(2b\mu/3)^{1/2}|{\bf d}(x_{F},T=0)|^{1/2}}
 {\pi T_{c}}\right]{\rm Tr}(\phi_{+}^{\dagger}\phi_{+})_{F}
 \nonumber \\ & &
 -\frac{7\zeta(3)}{8(\pi T_{c})^{2}}N(\mu/3)
 {\rm Tr}(\phi_{+}^{\dagger}\phi_{+}\phi_{+}^{\dagger}\phi_{+})_{F}\ ,
 \label{omegares1}
\end{eqnarray}
acts as a
counterterm to the coefficient of ${\rm Tr}(\phi_{+}^{\dagger}\phi_{+})_{F}$,
leading to $\Delta\Omega=0$ at $T=T_{c}$ \cite{BL}.  Then the sum of
$\Omega_{\rm loop}^{(0)}$ and $\Omega_{\rm res}$ reduces to the usual
Ginzburg-Landau part $\Omega_{0}$,
\begin{eqnarray}
 \Omega_{0}
 =N(\mu/3)
 \ln\left(\frac{T}{T_{c}}\right)
 {\rm Tr}(\phi_{+}^{\dagger}\phi_{+})_{F}
 +\frac{1}{2}\frac{7\zeta(3)}{8(\pi T_{c})^{2}}N(\mu/3)
 {\rm Tr}(\phi_{+}^{\dagger}\phi_{+}\phi_{+}^{\dagger}\phi_{+})_{F}
\nonumber \\
 \equiv
   \alpha^{+}{\rm Tr}(\phi_{+}^{\dagger}\phi_{+})_{F}
   +\beta^{+}_{2}{\rm Tr}[(\phi_{+}^{\dagger}\phi_{+})^{2}]_{F}\ .
 \label{alpha1}
\end{eqnarray}
From Eq.~(\ref{chi1}) we derive the value of $\chi$, (\ref{chi}).
Equation (\ref{alpha1}) indicates that $\beta_1^+= 0$,
and implies the coefficients in Eqs.\ (\ref{alpha})-(\ref{beta2}).

    The coefficients in Eq.\ (\ref{alpha1}) agree with those obtained by
Bailin and Love \cite{BL} for a BCS-type short-range pairing interaction and
for ${\tilde {\cal M}}=0$.  The agreement with the term of fourth order in
$\phi_{+}$ arises because the main contribution to the corresponding integral
in Eq.\ (\ref{loop00}) comes from the momentum region $||{\bf
q}|-\mu/3|\lesssim\kappa|{\bf d}(x_{F},T=0)|$, where the gap is almost flat.
The agreement for the second order term in $\phi_{+}$ is obvious.
On the other hand, the coefficient of ${\rm Tr}(\phi_{+}^{\dagger}
\phi_{+}{\tilde {\cal M}})_{F}$ in $\Omega_{\rm loop}^{(1)}$, as can be seen
from Eq.\ (\ref{chi1}), is different from $-(6/\mu)N(\mu/3)\ln(\Lambda/T_{c})$
with the ultraviolet cutoff $\Lambda$ as in a BCS superconductor, due to the
behavior $\phi_{+}({\bf q},T)\propto \sin({\bar g}y)|_{T=0}$ induced by the
long-range dynamically screened magnetic interactions.

    The difference in the ideal-gas free energy of the superfluid and normal
phases, $\Delta\Omega_{\rm ideal}$, Eq.\ (\ref{ideal}), gives rise to a term
proportional to ${\rm Tr}{\tilde{\cal M}}^{2}$, with coefficient given by Eq.\
(\ref{sigma}).  The color neutrality conditions, (\ref{neut}) and
(\ref{neutnd}), thus imply
\begin{equation}
  {\tilde\mu}_{ab}=
  -\frac{1}{3\mu}\ln\left(\frac{3T_{c}}{\mu}\right)
   [3d_{a}(x_{F},T)d_{b}^{*}(x_{F},T)-|{\bf d}(x_{F},T)|^{2}\delta_{ab}]\ ,
  \label{muaGL}
\end{equation}
in agreement with Eq.~(\ref{mua}) with $({\bf d}_{a})_{i}
=\delta_{is}d_{a}$. Comparing this expression to the $T=0$
result Eq.\ (\ref{muazero}), we find that ${\tilde\mu}_{ab}/|{\bf d}
(x_{F},T)|^{2}$ is identical at both $T=0$ and $T\simeq T_{c}$.
Expression (\ref{muaGL}) can also be derived from
the same analysis that yields Eq.~(\ref{muazero}) if we replace, in Eq.\
(\ref{nadif}), $\delta n$ as given by Eq.\ (\ref{deltan}) with
\begin{equation}
\delta n=
  V^{-1}\sum_{\bf q}
  \left\{\tanh\left(\frac{|{\bf q}|-\mu/3}{2T}\right)
  -\frac{|{\bf q}|-\mu/3}{E({\bf q})}
  \tanh\left(\frac{E({\bf q})}{2T}\right)\right\}\ .
\end{equation}
The equality in ${\tilde\mu}_{ab}/|{\bf d}(x_{F},T)|^{2}$ in the
$T=0$ and $T\simeq T_{c}$ cases arises from the fact that both have nearly the
same momentum dependence of the gap.

\section{Conclusion}
\label{sec:conclusion}

    In this paper we have laid out the Ginzburg-Landau structure of
superconducting quark matter.  Even in the homogeneous case considered here,
many questions remain.  First, electrical charge neutrality, in addition to
color neutrality, should be duly taken into account.  The system considered
here is composed of three flavor massless quarks in flavor equilibrium, as
characterized by Eq.\ (\ref{flaveq}).  Such matter, when normal, is
electrically neutral in itself.  Below $T_{c}$, however, it has nonzero net
electric charge, unless the order parameter is isotropic in flavor space.
This isotropy is retained by the color-flavor locked phase; however, in the
isoscalar channel, strange quarks remain gapless, leading to a deficit of the
total number of strange quarks relative to the total baryon number, and hence
to positive net charge.  This charge, in neutron star matter, would be
neutralized by charged leptons.

    It is straightforward to extend the formalism obtained in the absence of
leptons to the situation of electrical charge neutrality.  When the system is
in overall beta equilibrium, the quark chemical potentials, for each color
$a$, obey $\mu_{au}+\mu_e = \mu_{ad} = \mu_{as}$, where $\mu_e$ is the
electron chemical potential.  Then the quark chemical potential, $\mu_a$, for
color $a$, defined by
\begin{equation}
  \mu_a \equiv \mu_{ai} + q_i \mu_e\ ,
  \label{muae}
\end{equation}
is the same for all flavors $i$; here, $q_{i}$ is the electric charge of
the quark of flavor $i$.  This relation replaces condition (\ref{flaveq}).  As
a consequence of electrical charge neutrality, Eq.~(\ref{mu}) remains valid.
We keep the same definition (\ref{deltamu1}) of the ${\tilde\mu}_{aa}$.  The
matrix ${\tilde{\cal M}}$, given by Eq.\ (\ref{calM}), can now be rewritten as
a traceless matrix
\begin{equation}
  {\tilde{\cal M}}_{abij}=\delta_{ij}
   ({\tilde\mu}_{ab}-q_{i}\mu_{e}\delta_{ab})\ .
\end{equation}
Near $T_{c}$, $\Delta\Omega$ is still given by Eqs.\ (\ref{omega0d}) and
(\ref{omegacnd}), since the lepton pressure, which is of order $\mu_{e}^{4}$
$[\sim{\cal O}({\tilde\mu}_{ab}^{4})]$, is negligible compared with
$\Omega_{\rm CN}$ which is of order ${\mu^{2}\tilde\mu}_{ab}^{2}$.
(Up to leading order in ${\bf d}_{a}$, we obtain $\mu_{e}=-(\chi/2\sigma)
\sum_{ai}q_{i}[({\bf d}^{*}_{a})_{i}({\bf d}_{a})_{i}]_{F}+
{\cal O}(\mu_{e}^{3}/\mu^{2})$ from the electric neutrality condition,
$\partial\Delta\Omega/\partial\mu_{e}=0$, and Eq.\ (\ref{mua}) from
the color neutrality conditions (\ref{neut}) and (\ref{neutnd}).
Thus, $\mu_{e}\sim{\cal O}({\tilde\mu}_{ab})$.)  Since electricity does not
distinguish between colors, the presence of $\mu_{e}$ in the isoscalar
channel [$({\bf d}_{a})_{i}=\delta_{is}d_{a}$] leads to a Fermi momentum
mismatch between $u$ and $d$ quarks, and thus results in a uniform
suppression in the magnitudes of the gap and condensation energy;
in Eqs.\ (\ref{lambdaIS1}) and (\ref{EcondIS1}), $\beta_{1}$ is
replaced by $\beta_{1}+\Gamma/2$.  Accordingly, in the phase diagram
illustrated in Fig.\ 2, the isoscalar phase is shifted by $\Gamma/4$
in the $\beta_{1}$ direction and $-3\Gamma/4$ in the $\beta_{2}$ direction.
Note that the general flavor-antitriplet states belonging to this phase
can no longer be reduced to the isoscalar channel by flavor rotations since
the value of $\mu_{e}$ and the gap suppression depend on the the electric
charges of quarks involved in Cooper pairing.

    A second important problem is how the strange quark mass affects the quark
superfluidity.  A non-zero strange quark mass not only breaks invariance of
the grand-canonical Hamiltonian with respect to rotations in flavor space, but
it also necessitates the presence of a neutralizing gas of leptons even in the
normal phase.  The resulting $m_{s}$ dependence of the stability of
color-flavor locking over the isoscalar state was considered by several
authors (see, e.g., Refs.\ \cite{mass,henning}), who found that in contrast to
the case of the isoscalar channel, color-flavor locking is destabilized by the
Fermi momentum mismatch between $s$ quarks and $u$, $d$ quarks produced by a
nonzero value of $m_{s}$.

    It is instructive to consider the effect of $m_{s}$ in terms of the
general Ginzburg-Landau approach constructed here.  In the highly relativistic
regime ($m_{s}\ll\mu/3$), in addition to the terms in the massless limit,
$\Delta\Omega$ contains a term proportional to ${\rm Tr}(\phi_{+}^{\dagger}
\phi_{+}M^{2})_{F}$, where $M_{abij}=\delta_{ab}\delta_{ij}m_{i}$ with $m_{u}
=m_{d}=0$.  This term tends to suppress the pairing gap just like the term
proportional to ${\rm Tr}(\phi_{+}^{\dagger}\phi_{+}{\tilde{\cal M}})_{F}$
coming from color neutrality.  Note that because of color-flavor locking, the
resulting anisotropy of the condensate in flavor space acts to fix its
color orientation.  The transition temperature $T_{c}$ is also reduced
in the color-flavor locked state, leading to a $T_{c}$ in the color-flavor
locked state smaller than $T_{c}$ in the isoscalar channel, which is
independent of $m_{s}$.  The equilibrium phase diagram for the superfluid
transition, in the $\mu$ versus $T$ plane, is sensitive not only to the
effects of the color neutrality and the strong coupling, as stressed in the
present work, but also to effects of electric neutrality and finite strange
quark mass.

\acknowledgments

    We are grateful to T.~Hatsuda for helpful discussions.  Author KI would
like to acknowledge the hospitality of the Department of Physics of the
University of Illinois at Urbana-Champaign, and author GB the hospitality of
the Aspen Center for Physics during the course of this research.  This work
was supported in part by a Grant-in-Aid for Scientific Research provided by
the Ministry of Education, Science, and Culture of Japan through Grant No.\
10-03687, and in part by National Science Foundation Grant No.\ PHY98-00978.


\begin{references}
\bibitem{convention} Note that the convention for $\Delta$ here differs
from the usual condensed matter convention, $\Delta \sim
\langle\psi\psi\rangle$.
\bibitem{note2} Note that at fixed chemical potential, the number of the
quarks in a superfluid is larger than in a normal degenerate fluid, leading to
smaller chemical potential for a color with a larger gap in a color-singlet
quark superfluid.
\end{references}
\end{document}